\documentclass[preprint2]{aastex}

\usepackage{amsmath}
\usepackage{graphicx}
\usepackage{epsfig,psfrag,epic,eepic}
\usepackage{textcomp}
\usepackage{times}
\usepackage{longtable}
\usepackage{lscape}

\slugcomment{Submitted to the Astrophysical Journal}

\shorttitle{Multiplicity of M-dwarfs}
\shortauthors{Janson et al.}

\begin{document}

\title{The AstraLux Large M-dwarf Multiplicity Survey\altaffilmark{*}}

\author{Markus Janson\altaffilmark{1,2,5}, 
Felix Hormuth\altaffilmark{2}, 
Carolina Bergfors\altaffilmark{2}, 
Wolfgang Brandner\altaffilmark{2}, 
Stefan Hippler\altaffilmark{2}, 
Sebastian Daemgen\altaffilmark{3},
Natalia Kudryavtseva\altaffilmark{2},
Eva Schmalzl\altaffilmark{4},
Carolin Schnupp\altaffilmark{2},
Thomas Henning\altaffilmark{2}
}

\altaffiltext{*}{Based on observations collected at the European Southern Observatory, Chile, under observing programs 081.C-0314(A), 082.C-0053(A), and 084.C-0812(A), and on observations collected at the Centro Astron\'omico Hispano Alem\'an (CAHA) at Calar Alto, operated jointly by the Max-Planck Institute for Astronomy and the Instituto de Astrof\'isica de Andaluc\'ia (CSIC).}
\altaffiltext{1}{Princeton University, Princeton, USA}
\altaffiltext{2}{Max Planck Institute for Astronomy, Heidelberg, Germany}
\altaffiltext{3}{European Southern Observatories, Garching, Germany}
\altaffiltext{4}{Leiden Observatory, Leiden, Netherlands}
\altaffiltext{5}{Hubble fellow; \texttt{janson@astro.princeton.edu}}

\begin{abstract}\noindent
We present the results of an extensive high-resolution imaging survey of M-dwarf multiplicity using the Lucky Imaging technique. The survey made use of the AstraLux Norte camera at the Calar Alto 2.2m telescope and the AstraLux Sur camera at the ESO New Technology Telescope in order to cover nearly the full sky. In total, 761 stars were observed (701 M-type and 60 late K-type), among which 182 new and 37 previously known companions were detected in 205 systems. Most of the targets have been observed during two or more epochs, and could be confirmed as physical companions through common proper motion, often with orbital motion being confirmed in addition. After accounting for various bias effects, we find a total M-dwarf multiplicity fraction of $27 \pm 3$\% within the AstraLux detection range of 0.08--6\arcsec (semi-major axes of $\sim$3--227~AU at a median distance of 30~pc). We examine various statistical multiplicity properties within the sample, such as the trend of multiplicity fraction with stellar mass and the semi-major axis distribution. The results indicate that M-dwarfs are largely consistent with constituting an intermediate step in a continuous distribution from higher-mass stars down to brown dwarfs. Along with other observational results in the literature, this provides further indications that stars and brown dwarfs may share a common formation mechanism, rather than being distinct populations.  
\end{abstract}


\section{Introduction}
\label{s:intro}

The multiplicity of stars is an important characteristic for understanding how they form and evolve. The fraction of multiple stars has implications for typical architecture of stellar and planetary systems, monitoring their orbital characteristics yields information that would otherwise be unavailable, and their distribution in orbits and masses can yield insights into the formation process \citep[e.g.][]{Burgasser2007,Goodwin2007a}. One clear observational trend that has gradually emerged is that the fraction of stars that have stellar companions depends on the masses of those stars. For instance, within the population of Sun-like stars (FGK-type), less than half \citep[$\sim$46\%;][]{Raghavan2010} are multiple \citep[although a previous well-known survey gave a larger value of up to $\sim$67\%, see][]{Duquennoy1991}. By contrast, for more massive stars in the AB-type range, upwards of 80\% appear to be in multiples \citep{Shatsky2002,Kouwenhoven2007,Peter2012}. The trend of decreasing multiplicity from massive stars to Solar-mass stars also continues down to lower masses. In the M0--M6 spectral type range, multiplicity fractions of $\sim$26--42\% have been found \citep[e.g.][]{Fischer1992,Delfosse2004}. For even later-type objects in the VLM range (Very Low-Mass stars and brown dwarfs), the multiplicity frequencies are as low as $\sim$10--30\% \citep[e.g.][]{Bouy2003,Close2003,Gizis2003,Joergens2008}.

A challenging factor in previous surveys for multiplicity in low-mass stars has been limited sample sizes of $\sim$100 stars or less. Part of the reason for this is that M-stars are altogether more poorly characterized than higher-mass stars -- for instance, due to their low brightnesses, they are generally not included in the Hipparcos catalog \citep{Perryman1997}. Hence, well-characterized samples of M-dwarfs that are appropriate for statistical studies have been hard to come by. However, recently a number of efforts have been made in identifying and characterizing nearby M-dwarfs \citep[e.g.][]{Riaz2006,Reid2007,Lepine2011}, which has mitigated this situation. Following the publications of these studies, we have undertaken an extensive effort to study the multiplicity properties of nearby M-stars with high-resolution imaging. An intermediate study of 124 of these targets has been previously published \citep{Bergfors2010}. Here, we summarize the full sample of 761 individual stars and perform studies of statistical properties of various sub-samples.

A large sample of stars and a catalog of their binarity properties are relevant for a wide range of scientific purposes. For instance, the formation of VLM objects remains less well understood than for Sun-like stars. Although there are many indications that they may form as the low-mass tail of regular star formation \citep[e.g.][]{Luhman2005,Bourke2006}, there have also been alternative mechanisms proposed for the formation of some or all of these objects \citep[e.g.][]{Reipurth2001,Goodwin2007b,Basu2012}. For instance, \citet{Thies2007} argue that the different binarity properties of stars and VLM objects is an indication for separate formation mechanisms of the two populations. Sun-like stars have separations that peak broadly at $\sim$30~AU, and a rather uniform distribution of mass ratios \citep[e.g.][]{Duquennoy1991,Raghavan2010}, whereas VLM objects have separations that peak narrowly at smaller separations (3--10~AU), and show a preference to nearly equal masses of the components \citep[e.g.][]{Burgasser2007}. A key question, then, is whether the transition in binary properties from Sun-like stars to VLM objects proceeds smoothly over the intermediate range. A large M-dwarf sample is ideal for addressing this question. If any break or bimodality occurs over this range, this would imply that there are indeed two separate formation scenarios at play, whereas if the progression is fully continuous and monotonous, the opposite would be the natural conclusion. 

Besides statistical issues, the individual detections, confirmations, and observational parameters of binaries are highly useful. For instance, continued orbital monitoring can yield simultaneous brightnesses and masses of the components \citep[e.g.][]{Delfosse2000,Zapatero2004,Bouy2004,Liu2008}. Since our main sample from \citet{Riaz2006} is young, one of the implications of this is that the ages of individual systems can be inferred through isochronal analysis \citep[e.g.][]{Janson2007}. Conversely, for the youngest systems and lowest masses, the theoretical mass-luminosity relationships \citep[e.g.][]{Burrows1997,Chabrier2000}, which are increasingly uncertain under such conditions, can be better constrained \citep[e.g.][]{Hillenbrand2004,Konopacky2010}. 

The paper will be structured as follows: In Sect. \ref{s:obs}, we will describe the observations and data reduction procedure used for the survey. In Sect. \ref{s:photastr} the determination of the photometric and astrometric properties of each detected binary pair is described, along with the determination of the corresponding underlying physical parameters of the individual components. This is followed by a discussion of sub-sample selection for statistical purposes in Sect. \ref{s:sample}. Most of the binaries have been observed over two or more epochs, and the astrometric analysis of those cases is discussed in Sect. \ref{s:multep}. The statistical analysis is described in Sect. \ref{s:multfrac} for the multiplicity fraction, and Sect. \ref{s:physical} for the distribution of physical parameters. Finally, we summarize our results in Sect. \ref{s:summary}. Furthermore, individual notes for targets in the survey as well as a summary of background contaminants are provided in the Appendix.

\section{Observations and Data Reduction}
\label{s:obs}

For all observations in this survey, we have used AstraLux Norte \citep{Hormuth2008} at the Calar Alto 2.2m telescope in Spain, and AstraLux Sur \citep{Hippler2009} at the ESO/NTT 3.5m telescope on La Silla in Chile. The two AstraLux cameras are a near-identical twin pair of high-speed electron multiplying cameras developed for a wide range of high-resolution imaging purposes \citep[e.g.][]{Hormuth2007,Sicilia-Aguilar2008,Daemgen2009}. This is achieved through a technique known as "Lucky Imaging" \citep{Tubbs2002,Law2006}, where a very large number of frames ($\sim$10000) of very short integration time ($\sim$10~ms) are collected during an observation. In such a short time as 10~ms, each frame can be seen as capturing a frozen state of the wavefront distortion pattern imposed by the atmosphere. These states will be statistically distributed over a wide range of distortion severity, where in some cases, the wavefront will be particularly well preserved, so that the Strehl ratio will be high in these frames. By rejecting the majority of frames and using only the few percent of highest Strehl ratio, we can gain substantially in spatial resolution in this way, and achieve almost diffraction-limited images, at the cost of image depth. 

Our total sample consists of 761 unique stars, of which 124 were presented in \citet{Bergfors2010}. In this study, we present results for the full sample. In most cases where a companion candidate was found, multiple observations of the same system have been taken over a baseline that is useful for common proper motion tests, so that physical companionship can be stringently tested in those cases. The targets were chosen primarily from the \citet{Riaz2006} sample (569 stars) and additionally from the \citet{Reid2007} sample (198 stars), and are summarized in Table \ref{t:sample}. The spectroscopic distances in the table were adopted from \citet{Riaz2006} (but corrected for binarity for binaries detected with AstraLux, see Sect. \ref{s:sample}). The targets were selected directly from the above samples without additional constraints imposed, except for the AstraLux Sur targets, where only M-stars within 52~pc were observed. For statistical purposes, we have constructed homogenized sub-samples within the existing data. We will discuss various sub-samples that were selected for specific statistical investigations in Sect. \ref{s:sample}. With most stars being observed in two filters (SDSS $i^{\prime}$ and $z^{\prime}$) and many of the binaries being observed twice or more, our survey encompasses more than 1600 individual observations, each of $\sim$10--12 minute extent in telescope time. In total, this corresponds to approximately 300 hours of observations, which have been performed between January 2007 and February 2010.

The raw frames produced by the camera were reduced with the dedicated AstraLux pipeline \citep{Hormuth2008}, which makes the frame selection and recombination of frames, with a Drizzle algorithm to oversample the image and produce an output pixel scale of 23.3~mas/pixel for AstraLux Norte and 15.4~mas/pixel for AstraLux Sur. Flat field and bias corrections are applied prior to the drizzling. Individual frames were aligned based on the brightest pixel in the oversampled frames, and the 10\% best frames were selected for producing the final image. The pixel scale as well as the orientation of true North were monitored by observing astrometric references during the nights, in particular the Trapezium \citep{Kohler2008} whenever it was visible, and M15 \citep{Marel2002} at any other time. The representative uncertainties for these quantities, determined using the IRAF \textit{geomap} procedure, are $\sim$2~$\mu$as/pixel in pixel scale, which typically does not dominate the error in astrometric separation since the separations of the binaries in our survey are small, and $\sim$0.3$^{\rm o}$ in orientation angle, which often does dominate the error in position angle for the binary astrometry.

\begin{figure*}[p]
\centering
\includegraphics[width=16cm]{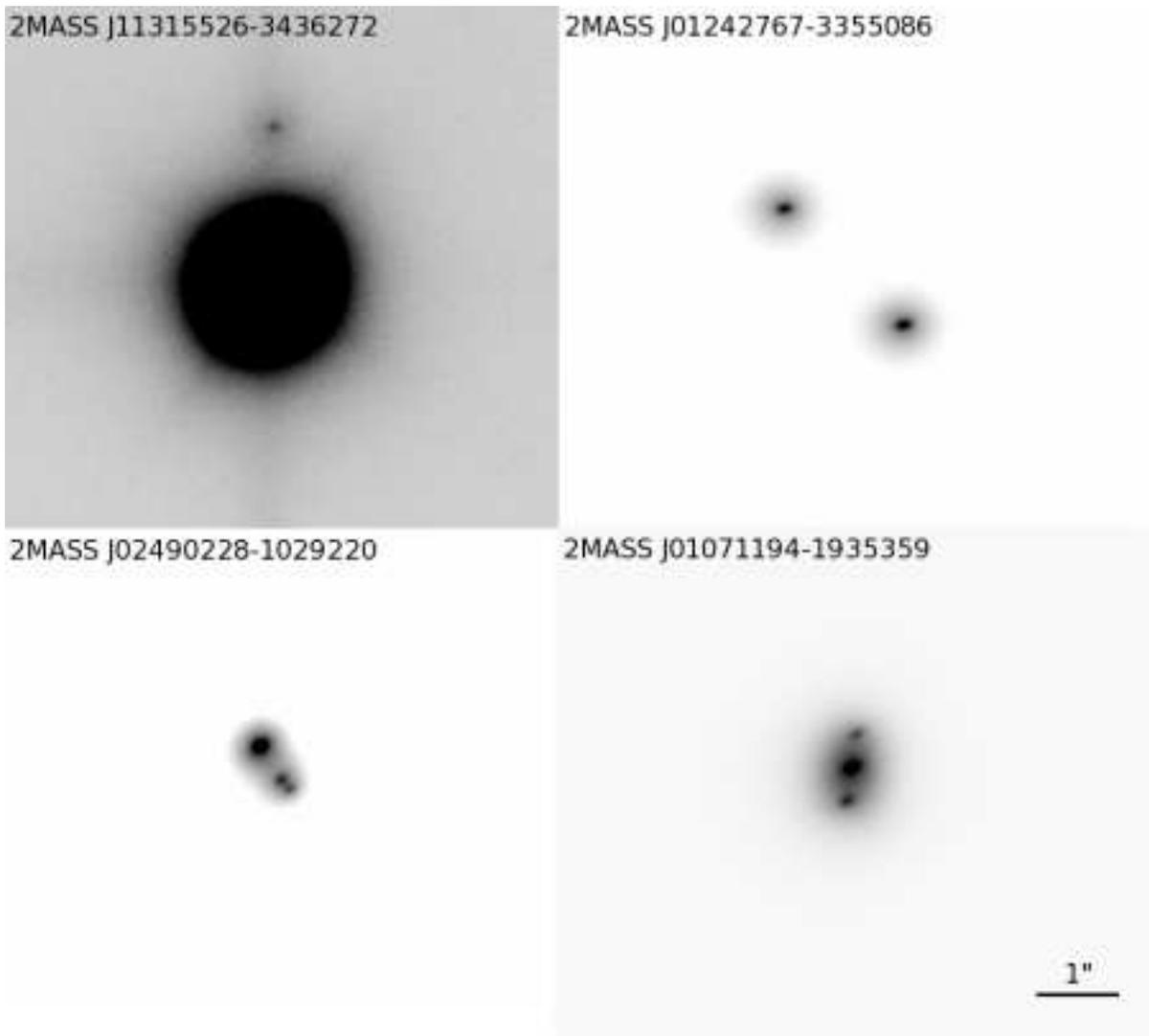}
\caption{Four examples of multiple systems imaged in the AstraLux survey. Top left: A high-contrast binary. Top right: A low-contrast binary. Bottom left: A triple system. Bottom right: A binary displaying the fake triple effect, common for close low-contrast binaries. All images are in $z^{\prime}$, and in each case North is up and East is to the left.}
\label{f:alux_im}
\end{figure*}

A sample of reduced images are shown in Fig. \ref{f:alux_im}. The separation and brightness contrast of each binary (as discussed in the following section) along with the AstraLux detection limit is shown in Fig. \ref{f:sep_dz}.

\section{Photometry and Astrometry}
\label{s:photastr}

The procedure for determining photometric and astrometric quantities was already described in \citet{Bergfors2010}, here we summarize that procedure. The relevant quantities to be determined are the magnitude differences between primary and secondary within each pair in each detected multiple system in $i^{\prime}$ and $z^{\prime}$ for photometry, and the projected separation and position angle of the components in each pair for astrometry. For close binaries ($\sim$1\arcsec\ or less), this was done with PSF fitting, using the reference PSF of a single star to iteratively create binary configurations that minimize the squared residuals of the fit. For wider binaries, we used Gaussian centroid fitting of the primary and secondary in order to yield the photometric and astrometric parameters. 

\begin{figure}[p]
\centering
\includegraphics[width=8cm]{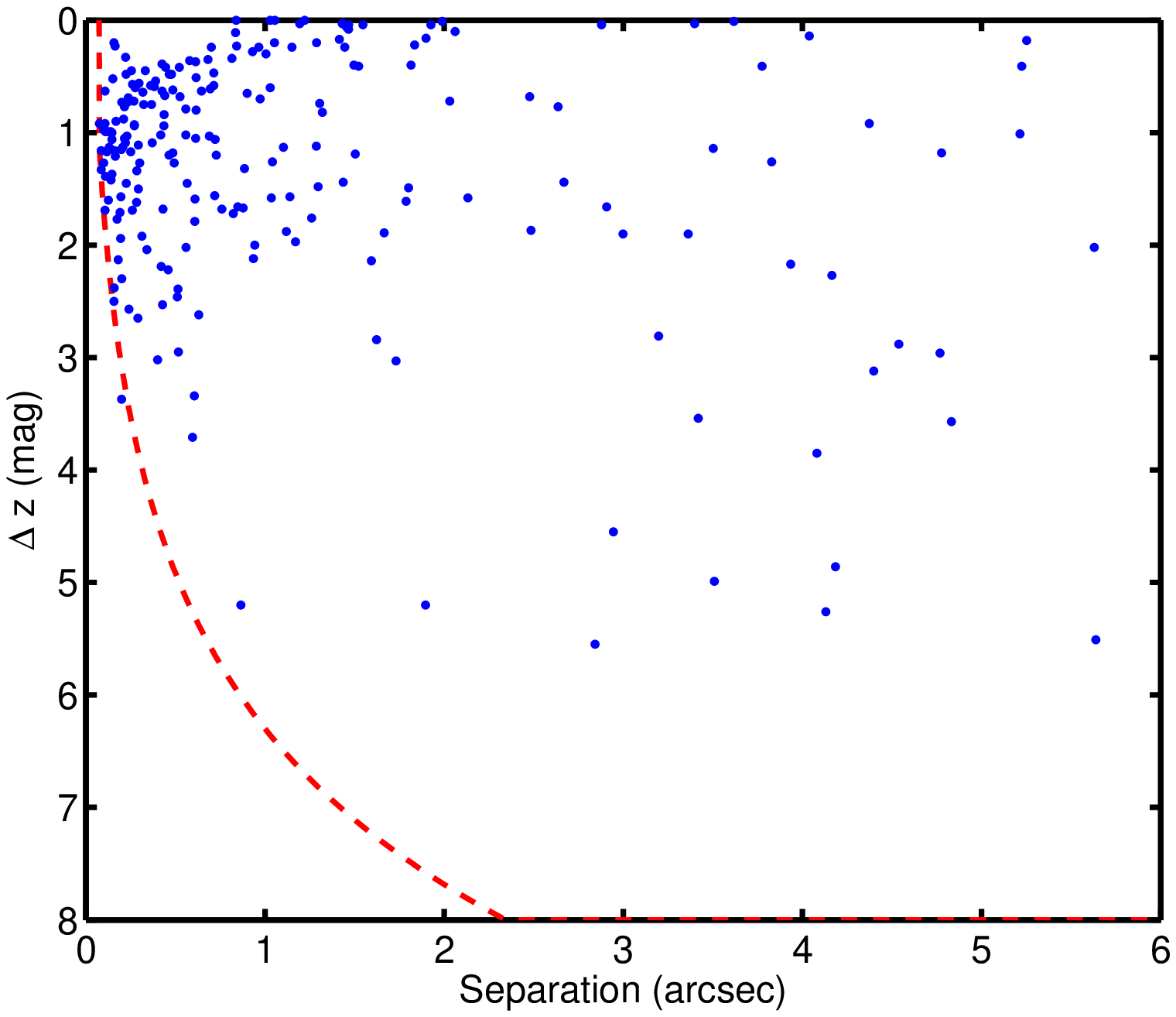}
\caption{The separation and brightness difference in $z^{\prime}$ for all binary pairs detected with AstraLux, plotted along with the typical AstraLux detection limit. The small zone in the upper left that shows a lack of objects is likely due to a bias that tends to systematically over-estimate the brightness difference for close pairs of nearly equal brightness.}
\label{f:sep_dz}
\end{figure}

As is normally the case in Lucky Imaging data, some close and near-equal component binaries display the "fake triple" effect, where a ghost tertiary appears at the same separation from the primary as the real secondary, but on the opposite side of it. In these cases, we use the procedure of \citet{Law2006thesis} to "de-triple" the system, determining the true flux ratio of the binary components $F_{\rm R}$ as

\begin{equation}
F_{\rm R} = \frac{2I_{13}}{I_{12}I_{13} + \sqrt{I^2_{12}I^2_{13} - 4I_{12}I_{13}}}
\end{equation}

where $I_{12} = F_1/F_2$ and $I_{13} = F_1/F_3$. When a fake triple occurs, or indeed more generally in any case of a close binary of nearly equal brightness, it is relevant to note that there can be a 180$^{\rm o}$ phase ambiguity, given the difficulty to determine which star is actually the primary. This is especially true for such late-type stars as these, which are sometimes quite variable, such that the binary components alternate between which is the brightest. On this topic, it should be noted that because the brightness difference between the two components is taken as the apparently brighter component minus the apparently fainter one, it is a positive definite quantity. Hence, the measured brightness difference on (e.g.) two equally bright components will not converge toward zero for a large number of measurements, since no measurement will be negative. This leads to a bias which exaggerates the brightness difference of close companions, and which most likely explains the apparent avoidance in the zone of very close binaries with almost equal-mass components (see Fig. \ref{f:sep_dz}). The final photometry and astrometry values of each detected binary are listed in Table \ref{t:photastr}, and in Table \ref{t:background} for the suspected or confirmed background contaminants..

In order to translate the integrated spectral type and relative photometry of each binary pair into individual spectral types and masses, we again follow the procedure in \citet{Bergfors2010}, which was developed in \citet{Daemgen2007} and is based on the magnitude-spectral type relationship in \citet{Kraus2007}. We base our final individual spectral types on the $z^{\prime}$-band, but we also make the same estimations on the basis of the $i^{\prime}$-band, which gives consistent spectral types to within $\pm$0.5 sub-classes (except in those cases that have been identified as background objects). The translation of individual spectral types to component masses also follows the relations given in \citet{Kraus2007}. The resulting physical quantities for the pairs are listed in Table \ref{t:physical}.

\section{Statistical Samples}
\label{s:sample}

Out of our total sample we construct two main sub-samples for the purpose of statistical analysis. This is done in order to remove statistical bias from selection effects and provide samples that are optimally suited for specific statistical studies. The first sample will be referred to as the Inclusive Sample (IS). This sample includes most of the stars observed, with some special cases excluded that are discussed toward the end of this section. However, this also includes 60 stars with estimated spectral types earlier than M-type (K7 and K5) which are additionally excluded in many of the statistical investigations. 

The second sample is a sub-sample of the first and is called the Constrained Sample (CS). This sample is constructed to provide a multiplicity fraction that is well-defined and bias free, following similar criteria to those in \citet{Bergfors2010}. In CS, we make a cut for significant X-ray flux. This provides a sample that is young \citep[less than $\sim$1~Gyr, see][]{Bergfors2010} and provides further confirmation that the selected targets are indeed nearby M-dwarfs rather than mis-classified distant objects, as can otherwise happen for these kinds of targets where a parallax is generally lacking. As in \citet{Bergfors2010}, we use X-ray counts and errors from ROSAT\footnote{Both the Bright Source Catalog IX/10A and the Faint Source extension IX/29} \citep{Voges1999} with 3.3$\sigma$ as the relevant significance criterion. In the case of binaries, this can lead to a selection effect due to the fact that the flux of the primary and secondary may add up to just above the significance criterion in cases where a single star would not have reached the criterion and thus not have been selected. We account for this in the same way as in \citet{Bergfors2010}, by noting that $L_{\rm X}/L_{\rm bol}$ is roughly constant as a function of spectral type, and calculating what the significance would have been for the primary component alone, had it been single. In those cases where the significance drops below the 3.3$\sigma$ criterion, the binary is removed from the CS. These cases are laballed `S' in the X-ray column of Table \ref{t:sample}.

We also adopt a 52~pc maximum distance cut-off for the CS, again analogously to \citet{Bergfors2010}. However, in addition, we also take into account a further bias that favors selection of binaries. Since most of the distances are determined spectroscopically under the assumption that the primary is single, a system that is in reality a binary will appear artificially too bright for the primary spectral type, and thus appear closer than it really is. We correct for this by adjusting all the distances for binaries that do not have parallax measurements and are not resolved in the 2MASS catalog \citep{Skrutskie2006}, which provided the brightness that the spectroscopic distance estimate was originally based on. Any binary that passes the 52~pc cut-off with the new distance estimate is removed from the CS. In the cases of both X-rays and distance, there can be stellar components inside of the resolution limit of AstraLux which could have similar effects as those corrected for here. To first order, one might expect that such effects should affect singles and multiples within the AstraLux range of 0.08\arcsec--6.0\arcsec\ equally and thus not introduce any significant bias in the multiplicity fraction, although it should also be noted that some bias could be introduced if there is a strong correlation between the presence of a close companion and the presence of wider companions.

The CS also only contains primary stars in the spectral type range of M0--M5. In contrast to  the analysis in \citet{Bergfors2010}, in this study we do not remove targets that are $>$6\arcsec\ companions to stars of earlier spectral type than M0. The reason for this is that we are concerned with multiplicity in the range of 0.08\arcsec--6.0\arcsec, and so any companions or lack of companions beyond these limits should not be allowed to have an impact on the multiplicity statistics.

In total, 761 targets observed by AstraLux are presented here, of which 205 are multiple within 0.08\arcsec--6.0\arcsec. The IS sample contains 751 targets with K-stars (693 without K-stars), of which 203 (191 without K-stars) are multiple. The CS sample contains 337 targets, where 85 are multiple. The majority of the multiple systems are binary, but there are 12 triple systems and 1 quadruple system in the IS, and 6 triple systems in the CS (all within 0.08\arcsec--6.0\arcsec). These numbers include only those detected candidates that are consistent with common proper motion with the primary and/or the expected color-magnitude relationship of a physical companion, and not those that have been rejected in this regard (see Sect. \ref{s:multep} for details on the companionship analysis).  

There are also 10 targets that are not included in either statistical sample. Seven of these were excluded because their multiplicity status is unclear: Since the resolution and contrast limits are not sharp cut-offs and since there is some variability in data quality, there inevitably exist limit-case tentative detections where binarity is plausible, but cannot be firmly established, for instance because the binary fitting does not converge, or there is a deemed risk of mix-up with some temporary PSF extension of non-astronomical origin. In our sample, these cases are J05323611-0523010, J15403153+1736554, J15530484+4457446, J21375368-0444383, J22581643-1104170, J23353509-0613302 and J23385413-1246184. Data under better conditions or with larger apertures are needed to confirm or reject those potential companions as astronomical objects. The remaining three cases that are not included in any of our sub-samples are binaries that contain a white dwarf. These systems, J03324345-0855391, J10162867-0520320 and J16291031+7804399, are discussed further in the individual notes.

\section{Multi-epoch Analysis}
\label{s:multep}

Most of the multiples observed with AstraLux have been detected in multiple epochs. Systems that were first discovered with AstraLux were generally observed over more than one epoch, and in addition, literature astrometry exists for some systems that had been previously discovered. In each case where multi-epoch astrometry exists, it becomes possible to stringently test physical companionship through common proper motion, as well as to examine whether significant orbital motion is taking place. Hence, we perform such analysis for all systems observed in more than one epoch. For each system, we acquire a proper motion for the primary from the literature, either from Hipparcos (where available), or otherwise from the PPMXL \citep{Roeser2010} or NOMAD \footnote{Available from http://www.nofs.navy.mil/nomad/ or VizieR I/297} catalogs. Likewise, whenever parallax values are available (see Table \ref{t:sample}), they are used in the analysis, in the remaining cases a parallax is calculated based on the spectroscopic distance estimate. Two epochs of astrometry are then chosen for each star, and the change in position is compared to the change that would be expected for an unrelated static background object. If the relative motion deviates by more than 3$\sigma$ from the background hypothesis, the binary is considered as a confirmed physical binary. If in addition there is a more than 3$\sigma$ deviation from exact co-motion, the binary is considered to display significant orbital motion, with some exceptions. These are the cases for which the estimated period is very large relative to the observed motion over the observational baseline. Although there can be a signficant excess of orbital motion over expectation if the orbit is close to periastron or if the distance to the system is considerably smaller than the mean value, there is also good reason for caution in the event that the astrometric errors have been underestimated in those cases. Hence, we regard such cases as unconfirmed with respect to orbital motion.

The alternative case where there is a less than 3$\sigma$ deviation from the background hypothesis is somewhat more complicated, because in that case there is an ambiguity between orbital motion and non-common proper motion -- in principle, there can be orbital motion that happens to bring the companion in the same direction as it would have moved relative to the primary, if it had been a static background source. There are a few cases in our sample where this issue is genuinely very complicated; these are discussed in the individual notes. However, in most cases, there are two clear groups of objects with non-significant common proper motion: those for which the expected motion is simply small compared to the astrometric errors, which we treat as unconfirmed with regards to companionship, and those that display a clear motion consistent with the background hypothesis, which we regard as being cases of background alignment. We summarize these results in table \ref{t:multep}. 

The vast majority of the systems examined do exhibit a common proper motion, which confirms that background contamination is rare and that most binaries in the total sample (including most of those that have only been observed in a single epoch) are real physically bound systems, as can also be statistically inferred simply from the fact that the frequency of detected companions increases rapidly with decreasing projected separation, rather than the other way around, as would be expected if background contaminants were dominant \citep{Bergfors2010}. Furthermore, the group of systems for which the color of the secondary is inconsistent with expectations for a physical companion provide an excellent match to the group of systems that are best consistent with the background hypothesis in the proper motion analysis. Hence, we conclude that false positives can be distinguished from real physical systems in our data with a high accuracy. All cases for which no common proper motion could be established (either due to only a single epoch being available, or due to insufficient motion with respect to the astrometric errors between the epochs) are listed in the individual notes in the Appendix. In order to quantify the risk of biasing the multiplicity statistics through mis-classification of background targets as physical companions and vice versa, we can note that only two of the systems that were classified through both their colors and their proper motion give conflicting conclusions between these two measurements (J03050976-3725058 and J06583980-2021526, see individual notes). Since 134 of the systems have been doubly tested in this regard, this means that 1.5\% of the systems are problematic, such that in the remaining 85 system that have not yet been doubly tested, we can expect one additional such system, giving three in total. Assuming that all these three systems are mis-classified gives, for instance, a potential impact on the multiplicity fraction among the 337 targets in the CS sample of approximately $\pm$1\%. Aside from being generous in this context given the fact that the two targets listed above are not even included in the CS, this is in any case smaller than the statistical error, as shown in Sect. \ref{s:multfrac}, and we can thus assume that the impact of this potential bias is negligible.

We do not make any attempts at more detailed orbital modeling even for systems with several epochs of observations -- additional AstraLux epochs are still being acquired, and a detailed orbital analysis based on those results will be the subject of a future publication.

\section{Multiplicity Fraction}
\label{s:multfrac}

\subsection{Total Multiplicity Fraction}

We calculate the total multiplicity fraction of M0--M5-stars between 0.08\arcsec\ and 6.0\arcsec\ based on our CS sample. The fraction of detected multiples in the CS is $85/337 = 25.2$\%. However, in order to get the actual multiplicity fraction in the range of 0.08\arcsec--6.0\arcsec, we have to account for the fact that the detectability is incomplete for faint companions at small separations. We do this by noting that detectability is essentially complete for all separations as long as $\Delta z^{\prime} \leq 1.2$~mag is fulfilled, and that it is complete for the full contrast range as long as $\rho \geq 1$\arcsec\ is fulfilled (see Fig. \ref{f:sep_dz}). We then assume that the distribution of $\Delta z^{\prime}$ is independent of separation, such that $N_{1,1}/N_{1,2} = N_{2,1}/N_{2,2}$ would hold true under full completeness, where $N_{1,1}$ is the number of binary pairs with $\Delta z^{\prime} \leq 1.2$~mag and $\rho < 1$\arcsec, $N_{1,2}$ is the number with $\Delta z^{\prime} \leq 1.2$~mag and $\rho \geq 1$\arcsec, and $N_{2,1}$ and $N_{2,2}$ are the corresponding cases with $\Delta z^{\prime} > 1.2$~mag. In other words, we assume that in the narrow-separation range of 0.08\arcsec--1.0\arcsec, the fraction of low-contrast companions ($\Delta z^{\prime} \leq 1.2$~mag) to high-contrast companions ($\Delta z^{\prime} \leq 1.2$~mag) is the same as is measured outside of 1\arcsec. Hence, the number of missed companions is the difference between the ideal $N_{2,1}$ that fulfills the equality and the actual measured number. In this context, it should be noted that the assumption of $\Delta z^{\prime}$ being independent of separation should be taken with some caution, as discussed in Sect. \ref{s:physical}.

In this way, we find that 7.4 companions are missing in the data, such that the actual multiplicity fraction is $(85+7.4)/337 = 27.4 \pm 3.1$\%, where the error is based on Poissonian statistics. There is a remaining incompleteness to low-mass brown dwarfs; although we are sensitive to L0-type companions out to 52~pc, later-type companions fall below the sensitivity limits for the distant part of the sample. However, given that such companions are rare even for closer systems (only three $>$L0 companions exist in the entire observed sample, see Table \ref{t:physical}), this is most likely a marginal effect. Hence, in summary, we conclude that the multiplicity fraction of M0--M5-stars within 0.08\arcsec--6.0\arcsec\ is $27 \pm 3$\%.

As mentioned above, our observations and statistical studies are focused on acquiring multiplicity statistics within the separation range covered by AstraLux of 0.08\arcsec--6.0\arcsec. However, in order to put this discussion in a broader context, we will briefly discuss the implications for the total multiplicity fraction, irrespective of separation. We do this by comparing the result to the multiplicity study by \citet{Fischer1992} (abbreviated as FM92), which studies the full separation range by combining radial velocity and imaging data. Our limits in angular separation at the median distance of the sample of 30~pc correspond to projected separations of 2.4~AU to 180~AU. We translate this into statistically relevant physical separations by applying the FM92 correction factor of 1.26, yielding physical limits of 3~AU to 227~AU. Out of the 37 companions detected in FM92, 14 are either inside of 3~AU or outside of 227~AU. Henceforth, we will assume that the completeness within 3--227~AU is approximately the same as outside of this range in FM92; the estimated completeness values in FM92 largely support such an assumption, although the values for very large separation are highly uncertain. Under this assumption, we can hypothesize that in addition to the 85 companions that we detect within 3--227~AU in our CS sample, there additionally exist $85*14/37 = 32.2$ companions outside of these limits. Some of these will exist in systems that appear single in the AstraLux images, and thus form an additional multiple system, whereas others will exist in systems the are already binary within 3--227~AU, and thus not affect the total multiplicity fraction (but only the higher-order fraction). To which extent these additional companions affect the multiplicity statistics thus depends on to which extent close or wide companions are correlated or anti-correlated with companions in the intermediate separation range. For the purpose of this discussion, we assume that there is no correlation, such that exactly 27.4\% of the additional components occur in systems that had already been identified as multiple. This leaves 23.4 companions that contribute to the total multiplicity fraction over all separations, which consequently comes out to 34.4\%. By comparison, the total multiplicity fraction in FM92 is given as $42 \pm 9$\%. Alternatively, we may recalculate the FM92 multiplicity within 3--227~AU under the same assumptions, by removing multiplicity components that are outside of these ranges. This gives a fraction of 28\% within 3--227~AU in FM92. Hence, our data implies a slightly lower multiplicity fraction than FM92, but the values are consistent within the errors (under the given assumptions).

\subsection{Multiplicity Dependence on Spectral Type}

After having examined the multiplicity fraction of the full population of M0--M5-stars, we turn to studying the dependence of multiplicity fraction on spectral type within the sample. It is already well known that this fraction increases globally with increasing primary mass, since the multiplicity fraction is higher for Sun-like stars at 46--67\% \citep{Duquennoy1991,Raghavan2010}, yet higher for yet more massive stars at $>$70\% \citep[e.g.][]{Shatsky2002,Mason2009}, and lower for brown dwarfs at 10--30\% \citep[e.g.][]{Burgasser2003,Joergens2008}. Given our large sample and relatively large range in stellar masses, we can check how the multiplicity fraction evolves as function of primary mass self-consistently within the sample. In other words, if there are any discontinuities along the range as opposed to a smooth decline toward smaller masses, these could potentially be noticeable in the data. We use the IS M-dwarf sample for this purpose. 

\begin{figure}[p]
\centering
\includegraphics[width=8cm]{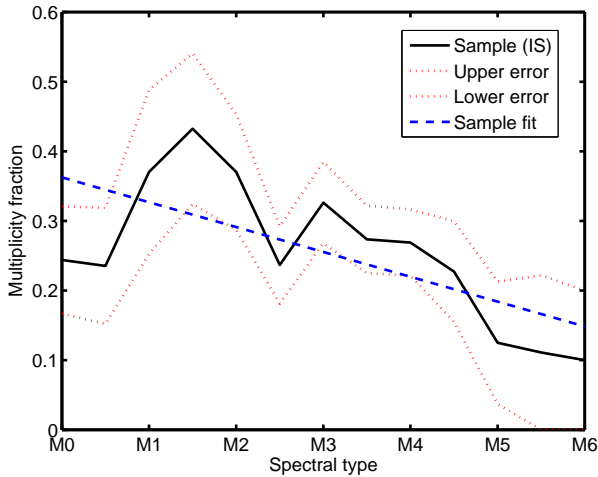}
\caption{Multiplicity fraction as a function of primary spectral type. The data are consistent with a uniformly increasing trend with stellar mass.}
\label{f:multfrac}
\end{figure}

A plot of the multiplicity fraction as function of spectral type is shown in Fig. \ref{f:multfrac}. The distribution is consistent with a uniform trend, and with random scatter from Poisson errors around the trend. For instance, although the fraction at spectral type M0 is lower than the fraction at M2, this is not statistically significant. We show this by making a linear fit to the data. The best fit (shown in Fig. \ref{f:multfrac}) gives a $\chi^2$ of 10.1, corresponding to a probability that it matches the underlying distribution of 52.6\%. Hence, a monotonous trend is fully plausible given the data, and there is no valid motivation to infer any sub-structure in the general trend. Furthermore, the slope of the line is negative with increasing sub-class (i.e., decreasing mass), with a coefficient of $-0.025 \pm 0.013$. Thus, at a marginal significance of 1.9$\sigma$, we can confirm the trend of multiplicity function with stellar mass, even within just the M-star spectral type range.

\section{Physical Distributions}
\label{s:physical}

\subsection{Distribution in Mass Ratio}

The mass ratios of binaries among Sun-like stars and brown dwarfs have quite different distributions \citep[e.g.][]{Burgasser2007,Raghavan2010}, where the mass ratios among Sun-like stars are quite uniformly distributed, while the components of brown dwarf binaries appear to be preferentially nearly equal in mass. It is thus interesting to study to whether M-dwarfs constitute a transitional regime between these two distributions, and if so, whether the transition is continuous or discrete. In order to demonstrate that this is a more complicated issue to address than it might seem at first, we first discuss the case of the IS sample.

\begin{figure}[p]
\centering
\includegraphics[width=8cm]{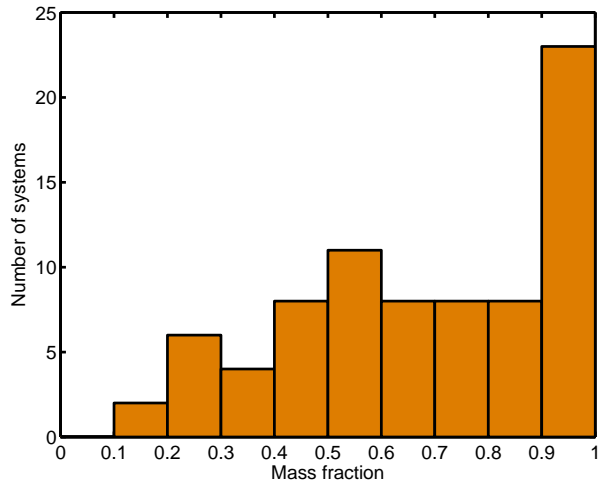}
\caption{Histogram of mass ratios for the IS sample. The peak at near-equal masses is due to an observational bias, as explained in the text.}
\label{f:massfrac_bias}
\end{figure}

\begin{figure}[p]
\centering
\includegraphics[width=8cm]{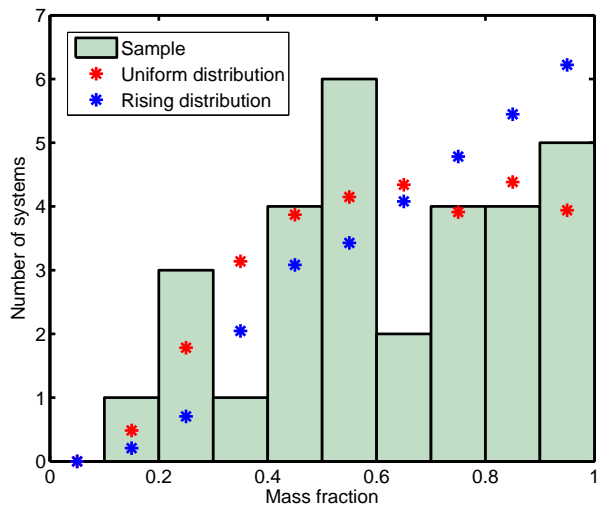}
\caption{Histogram of mass ratios for the CS sample. Also plotted are the bias-corrected simulated distributions for the uniform and rising cases, normalized to the same number of total systems. The sample is better consistent with uniform distribution than a rising one.}
\label{f:massfrac_real}
\end{figure}

A histogram of the mass ratios in the IS sample is shown in Fig. \ref{f:massfrac_bias}. Here the mass ratio is defined as $q_{\rm m} = m_{\rm b}/m_{\rm a}$, where $m_{\rm a}$ is the primary mass and $m_{\rm b}$ is the secondary mass. The sample shown in the figure only includes binaries with separations larger than 1\arcsec, since smaller separations are subject both to bias due to the limited detectability of high-contrast systems in that range, as well as the high uncertainty in determining flux ratios of close binaries. The histogram shows a very pronounced peak at mass ratios near unity, in addition to a population with a more homogenous distribution. However, the peak is not physically real, as we can see if we plot the same distribution for the CS sample, shown in Fig. \ref{f:massfrac_real}. Again, this analysis contains only binaries with separations greater than 1\arcsec. For the CS sample, the distribution is much more homogenous, with no sharp peak near equal masses. The stark difference between the two samples can be easily understood by considering the biases that are involved. As we mentioned in Sect. \ref{s:sample}, in the CS sample we have de-selected binaries that were positively selected for during observation preparations by appearing to be closer or brighter in X-rays than they would have been if they had been single. This effect is by far the strongest in binaries where the components have about equal brightness, and thus about equal mass. Hence, even though this is globally a subtle effect, in that the number of binaries that are positively selected for is rather small compared to the total number of binaries and single stars in the survey, it is the case that in every event where it happens, the contaminating binary has a near-unity mass ratio and thus the effect becomes significant in the particular case of mass ratio investigations.

For this reason, we will concern ourselves only with the CS distribution for the rest of the mass ratio discussion. In order to examine this distribution in a bias-free manner, we need to take some observational effects into account. As a first step, we select only pairs with $>$1\arcsec\ separations, as described above. In addition, we need to take into account the fact that the mass ratio in systems containing a brown dwarf candidate is highly uncertain, since brown dwarfs cool continuously after formation. Thus, the mass-luminosity relation depends on the highly uncertain age in those cases. We therefore remove all systems containing any component with a spectral type of L0 or later from our further analysis on the mass ratio distribution. However, this also introduces a bias (by preferentially removing systems with low mass ratios) that needs to be accounted for. We do this by generating simulated populations and subjecting them to the same process, as described in the following.

As mentioned above, the most interesting scientific question regarding the mass ratio distribution concerns whether this distribution is best described as uniform, as for the Sun-like primary population, or as rising toward near-equal masses, as in the case of the brown dwarf population and as some star formation simulations predict should be the case also for low-mass stars \citep{Bate2012}. We therefore generate simulated populations that follow these types of distribution. For each primary mass in our sample, 100 random mass ratios are generated. These are either uniformly distributed between 0 and 1 (to represent the uniform case), or distributed between 0 and 1 in such a way that the probability is directly proportional to the mass ratio (to represent the rising case). In every case where $q_{\rm m}*m_{\rm a} < 0.08$~$M_{\rm sun}$, the system is removed from the analysis. The full population of all mass ratios for all stars is then compared to our observed sample using a Kolmogorov-Smirnov test (K-S test). Furthermore, this process is repeated 100 times performing the same test, in order to ensure that the result is statistically relevant. When assessing the probability that the two compared samples are drawn from the same distribution, we use the mean probability from the 100 tests. We find that the probability that our observed sample is consistent with a uniform distribution is 63.7\%, but the probability for the rising distribution is only 8.9\%. The simulated distributions are shown in Fig. \ref{f:massfrac_real}.

We also make the same comparison of a uniform and rising sample to the \citet{Bate2012} sample of simulated star formation. In this case, we do not remove brown dwarfs since all samples are simulated and have no observational bias. As expected, the \citet{Bate2012} sample provides a better match to the rising distribution with a probability of 39.4\%, compared to the uniform distribution which has a probability of 11.1\%. Hence, it follows that the mass distribution of our sample of M-stars is better described by a Sun-like uniform distribution, than by a distribution that moderately favors near-equal masses as in the \citet{Bate2012} simulations, which one might have expected for an intermediate continuous population between Sun-like stars and brown dwarfs. This could imply that a sharp gradient or discontinuity occurs somewhere near the star/BD boundary, which would be difficult to consolidate with a common formation mechanism across this mass range. However, this issue should be considered with some caution, given the uncertainties in determining masses for brown dwarfs, as mentioned above. Also, it could for instance be the case that the mass ratio distribution depends on semi-major axis \citep[e.g.][]{Delfosse2004,Bate2009}. Here we have studied a relatively wide population of binaries, since we removed $<1$\arcsec\ binaries from the comparison. The mass ratio distribution of Sun-like stars is similarly dominated by wide binaries since their semi-major axis distribution peaks at such separations. By contrast, the brown dwarf population is dominated by small semi-major axes. A future specific study of M-dwarf multiplicity at small separations would be interesting in this regard (as well as an extensive survey of BD binarity at wide separations).

\subsection{Distribution in Semi-Major Axis}

\begin{figure}[p]
\centering
\includegraphics[width=8cm]{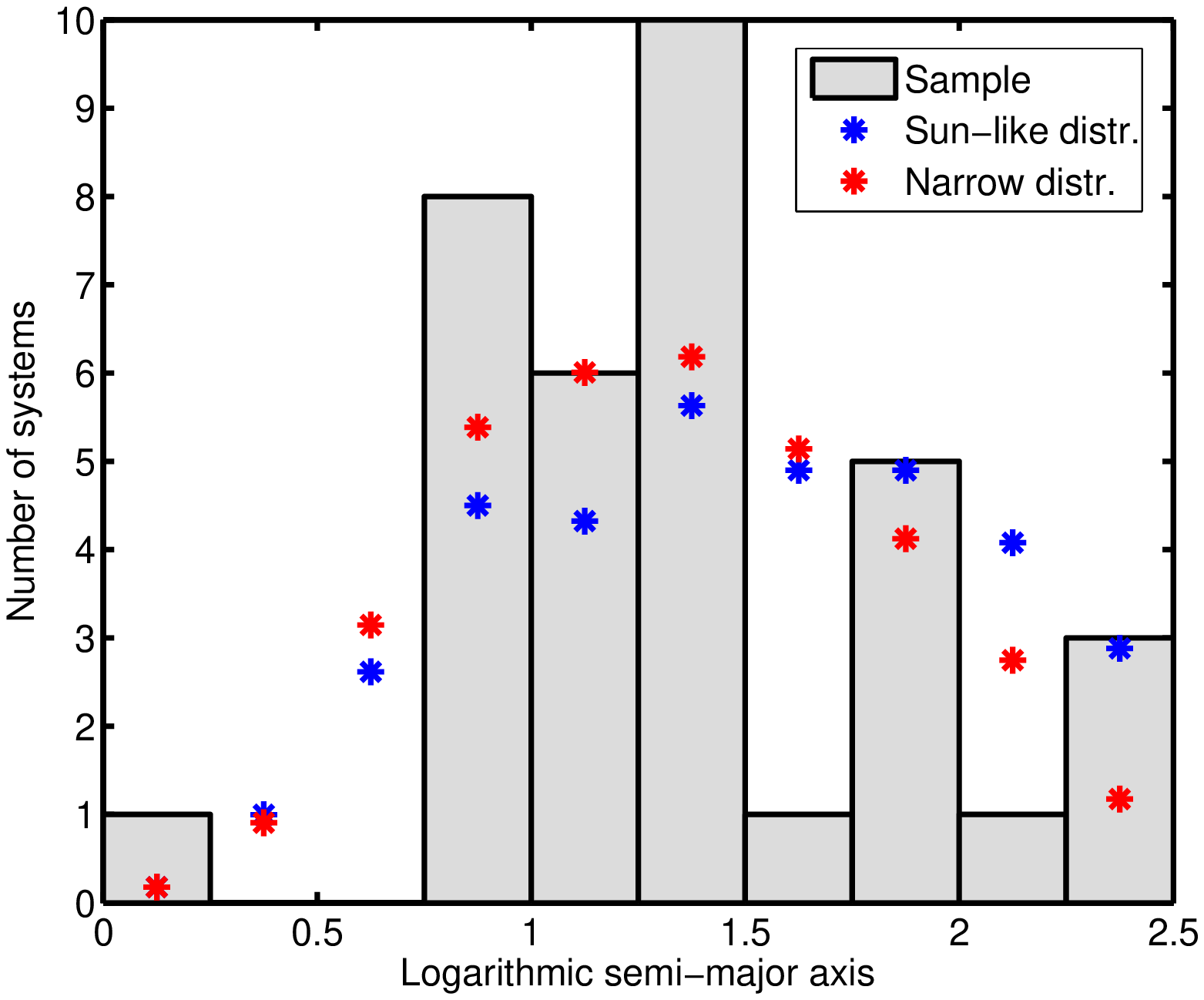}
\caption{Histogram of semi-major axes for the CS sample. Also plotted is the bias-corrected simulated distribution for Sun-like stars, as well as a narrower distribution peaking at smaller separations. The sample is inconsistent with a Sun-like distribution, but better consistent with the narrower one.}
\label{f:semimaj}
\end{figure}

Since our survey has a typical inner semi-major axis cut-off at 3~AU which is somewhat comparable to the most interesting region of distinction between the stellar and brown dwarf multiplicity populations at $\sim$10~AU, there is reason to proceed with care when determining this distribution in our sample. First of all, we use the CS sample for this purpose, in order to avoid distant targets that might skew the distribution. Also, we make a cut in flux ratio, keeping only binaries with $\Delta z^{\prime} < 1.2$~mag, since there is a gradual loss of sensitivity to fainter companions inside of 1\arcsec, which could affect the apparent distribution. Furthermore, we include only stars with primary mass $>0.2$~$M_{\rm sun}$, since lower-mass stars are not fully complete out to 52~pc and thus have smaller average distances than the full sample, which again could skew the distribution. These cuts leaves a sub-sample of a rather modest size, whose semi-major axis distribution is plotted in Fig. \ref{f:semimaj}. 

The distribution appears to peak at smaller semi-major axes and be narrower than the corresponding distribution for Sun-like stars. In order to test this, we simulate binary populations with a Gaussian distribution in logarithmic semi-major axis space, with $\mu_{\rm sim} = 1.64$ and $\sigma_{\rm sim} = 1.52$ from \citet{Raghavan2010}. For each star in our statistical sub-sample with a distance $d$, we generate 100 random semi-major axis values $a$ following the above Gaussian distribution. In every case where $a/d/1.26 < 0.08$\arcsec\ or $a/d/1.26 > 6.0$\arcsec, the simulated pair is removed from the analysis, in order to reproduce the existing bias in the observed population. The simulated population is then compared with the observed population using a K-S test, and this test is repeated 100 times in the same way as described in the previous section. This gives a 5.8\% probability that the populations are drawn from the same distribution. As a reference case, we also repeat the same procedure for an example distribution with a lower mean and a narrower width ($\mu_{\rm sim} = 1.2$ and $\sigma_{\rm sim} = 0.8$), which gives a significantly higher probability of 55.5\%. Hence, the results imply that M-dwarfs indeed have a semi-major axis distribution that is narrower and peaks at smaller values than for Sun-like stars. At the same time, the distribution seems to peak at larger separations than brown dwarfs, since it starts to drop already at $\sim$10~AU, which is probably not a bias-dominated trend, given that the median inner limit of detectability is 3~AU in our sample. This apparently intermediate distribution between Sun-like stars and brown dwarfs could be interpreted as another piece of evidence in favor of a common formation mechanism.

\section{Summary and Conclusions}
\label{s:summary}

In this article, we have presented results for a high-resolution imaging survey of 761 low-mass stars with the AstraLux Norte and AstraLux Sur cameras, using the Lucky Imaging technique. We have detected 219 companions in 205 multiple systems, of which 182 companions in 171 systems were previously undiscovered. Our census of multiplicity in these young nearby M-stars is potentially useful for a wide range of future applications, such as calibration of theoretical isochrones for low-mass stars and determination of orbital distributions through continued orbital monitoring. The sample should also be useful for future planet-search projects, such as high-contrast imaging surveys, which typically seek to avoid close binaries as targets, or astrometry surveys with planned instruments such as GRAVITY \citep{Eisenhauer2011}, for which close binarity is highly desirable for astrometric referencing purposes.

Our sample also allows us to study the population statistics of M-dwarf binarity. We find that the multiplicity fraction of M0--M5-type stars between separations of 0.08\arcsec\ and 6\arcsec\ is $27 \pm 3$\%. If we assume that close and wide binarity is uncorrelated with intermediate-range binarity, this implies a total multiplicity fraction in the range of $\sim$34\%, slightly lower than reported by \citep{Fischer1992} but consistent within the errors. Furthermore, we find that the multiplicity fraction as function of primary mass in our sample is consistent with a smooth increase in multiplicity with increasing mass across the sample, as might be expected from the fact that the multiplicity is known to be higher for Sun-like stars and yet higher for high-mass stars, and lower for brown dwarfs. The mass ratios of the binary components in our sample on the other hand appear to be fully consistent with a uniform distribution and no preference toward near-equal masses. This is consistent with an equal distribution to Sun-like stars, which is distinct from the distribution of the brown dwarf population and may indicate a rather abrupt change in properties in the vicinity of the star-BD boundary. The semi-major axis distribution is however again consistent with a smooth intermediate stage between stars and brown dwarfs. Hence, the different multiplicity properties give conflicting information about whether or not the transition from stars to brown dwarfs is smooth, and thus whether or not they are best interpreted as sharing a common formation mechanism. On balance, given the two pieces of evidence in favor and the single one against, we consider a common mechanism to be more likely. We emphasize that no final conclusions can be based on these distributions alone, but that they should be interpreted as pieces of evidence in a larger picture, along with other lines of evidence from observations and simulations. Of course, it could also be the case that several different mechanisms are involved in the formation of brown dwarfs and very low-mass stars in overlapping mass ranges \citep[e.g.][]{Dupuy2011,Riaz2012}. 

Most of the binaries and higher-order multiples in our sample have been observed at two or more epochs, which allows to confirm common proper motion (as is expected for physical pairs but not for background contaminants) and to get a first handle on orbital motion. The vast majority of point-sources in the AstraLux images are real physical companions, and the large difference in colors between companions and typical background stars allow us to distinguish between these two cases even for the targets that have only been observed in one epoch. Many of the targets show highly significant orbital motion, and could have their complete orbital characteristics determined through further monitoring over a longer baseline. 

An even more stringent and fully encompassing census of the multiplicity characteristics of this sample could be acquired through monitoring with radial velocity. This would allow to discover binaries on smaller orbits, as well as to determine accurate mass ratios and various orbital elements of the binaries. In this way, many biases could be removed and a much larger fraction of the sample used for a wider range of statistical applications. We are already studying a subset of targets with this technique, and future spectrographs that are well suited for radial velocity at near-infrared wavelengths, such as CARMENES \citep{Quirrenbach2010} should be able to further enhance such studies.

\acknowledgements
We thank all the staff at the Calar Alto and La Silla observatories for their support, and the referee, V.J.S. B\'ejar, for his useful report. This study made use of the CDS services SIMBAD and VizieR, as well as the SAO/NASA ADS service. Support for this work was provided by NASA through Hubble Fellowship grant HF-51290.01 awarded by the Space Telescope Science Institute, which is operated by the Association of Universities for Research in Astronomy, Inc., for NASA, under contract NAS 5-26555.

\appendix
\section{Notes on individual binaries and multiple systems}
Special remarks on individual systems are summarized below.

\textbf{J00080642+4757025}
In addition to the companion detected with AstraLux, J00080642+4757025 is a close spectroscopic binary with a 4.4 day period \citep{Shkolnik2010}, hence the system is actually a triple.

\textbf{J00150240-7250326}
The AstraLux PSF was strongly asymmetric for this target, hence in addition to the usual PSF fitting, we also performed a customized aperture photometry scheme in order to ensure that the PSF fitting did not provide a biased result. We chose a circular aperture with a 3 pixel radius centered on the primary and the companion respectively. In order to estimate the influence of the primary PSF at the location of the secondary, we adopted the mean of aperture fluxes in two locations at the same separation as the secondary, azimuthally separated from it by $\sim$15 pixels in each direction. This mean value was subtracted from the aperture flux of the secondary, thus removing the influence of the stellar PSF. The results are well consistent with the PSF fitting values and the method seems more sensible for this particular type of application, hence we adopt the aperture photometry values for the differential photometry, and let the errors be represented by the scatter between PSF fitting results and the aperture photometry. Note that the photometry values presented here should replace those in \citet{Bergfors2010}.

\textbf{J00193931+1951050}
J00193931+1951050 and J00194303+1951117 are separated by only 53\arcsec, have very similar proper motions \citep[e.g.][]{Roeser2010} and have similar estimated spectroscopic distances (23 and 21 pc, respectively) in \citet{Riaz2006}, hence it is highly likely that they form a physical pair.

\textbf{J00250428-3646176}
Like J00150240-7250326, this stellar binary also had an asymmetric PSF, and we therefore calculated the photometry in the same way as for that case, see the individual note for J00150240-7250326. Note that the photometry values presented here should replace those in \citet{Bergfors2010}.

\textbf{J00424820+3532554}
This system is noted as being triple in \citet{Tokovinin2006}, including a close spectroscopic binary pair with approximately 1.3 mas separation and a likely wider companion at 11\arcsec\ separation. Due to these separations however, it is counted as single for our purposes.

\textbf{J00560596+4153282}
The star has a likely companion at 13\arcsec\ with similar proper motion \citep[e.g.][]{Roeser2010} seen in 2MASS, but appears single closer in, in the AstraLux field of view.

\textbf{J01034210+4051158}
Aside from the companion seen in the AstraLux images, there is an additional possible wide companion at 26\arcsec\ separation, as noted in the Washington Double Star (WDS) catalog \citep{Mason2001}. The wide components have very similar proper motions \citep[e.g.][]{Roeser2010}, so actual companionship seems probable.

\textbf{J01093874-0710497}
This binary displays clear common proper motion, but the residual astrometry does not seem to follow a simple trajectory consistent with only orbital motion of the two visible components. In addition, the brightness difference is larger than the spectral type difference would imply at a common distance. This could indicate that the A component is itself an unresolved binary.

\textbf{J01132817-3821024} 
The primary star in the system is a known eclipsing binary with a period of 0.4455 days \citep{Parihar2009}. Since the angular separation is inside of our inner cut-off, we count the system as a binary here for the purpose of multiplicity statistics, although it is really a triple system. To estimate the mass of the primary for the purpose of deriving the mass ratio to the tertiary component detected in our images, we use the sum of the masses corresponding to the individual spectral types, M1 and M3, derived for the eclipsing binary.

\textbf{J01154885+4702259} 
In addition to the companion detected with AstraLux, it can be noted that J01154885+4702259 has a very similar proper motion to the star J01155017+4702023 at just 27\arcsec\ separation \citep{Roeser2010}, which is itself a close binary \citep{Law2008}. Thus, this is a very likely quadruple system.

\textbf{J01452133-3957204}
J01452133-3957204 has only been observed in one epoch, but the color and brightness of the detected companion are consistent with expectation, hence we count it as an unconfirmed binary.

\textbf{J01473248+3453528} 
This star has a possible wide brown dwarf companion at 43\arcsec\ separation, according to \citet{Casewell2008}.

\textbf{J01564714-0021127} 
This is an eclipsing binary with a 0.5 day period \citep{Chen2006}, but it is single in the AstraLux images.

\textbf{J02002975-0239579}
The companion to J02002975-0239579 has yet to be tested for common proper motion, but since its color and brightness are consistent with expectation and its separation is small ($\sim$0.32\arcsec), it is counted as a physical companion here.

\textbf{J02070176-4406380}
J02070176-4406380 and J02070198-4406444 are separated by only 7\arcsec\ and have similar estimated spectroscopic distances (13 and 21 pc, respectively) in \citet{Riaz2006}, hence it is likely that they form a physical pair.

\textbf{J02155892-0929121}
This likely triple system has only been observed in one epoch, hence it remains unconfirmed. The brightnesses and colors of all components are however consistent with expectation for a physical triple.

\textbf{J02165488-2322133}
The companion to J02165488-2322133 has not yet been confirmed to share a common proper motion, but the color and brightness of it are well consistent with a physical companion, hence it is counted as such in the statistics.

\textbf{J02335984-1811525}
There is insufficient motion between the two epochs of observation of J02335984-1811525 to establish common proper motion of the binary, but the components are nearly equal brightness and color, and reside at a relatively small separation of $\sim$0.85\arcsec\, hence it is almost certainly a physical pair.

\textbf{J02411909-5725185 }
J02411909-5725185 is yet to be confirmed as a common proper motion binary, but the color and brightness of the companion are well consistent with it being physically bound, hence it is counted as a binary in the statistics.

\textbf{J02411510-0432177}
This star has been classified as a possible T Tau star in the Taurus-Auriga region by \citet{Li1998}.

\textbf{J02451431-4344102}
J02451431-4344102 is a member of an approximately equal-brightness binary with $\sim$50\arcsec\ separation \citep{Mason2001}, in addition to being a close binary discovered with AstraLux.

\textbf{J02490228-1029220}
The system is precisely at the 3$\sigma$ threshold for statistically significant exclusion of the background star hypothesis. By chance, the AB astrometry is just below the criterion (2.7$\sigma$) and BC is just above (3.1$\sigma$), so in principle the BC pair is inconsistent with chance alignment, if the pair does share a common proper motion with A. However, the proper motion of the system has of course been determined based on A, so until the AB physical connection has been demonstrated to the required accuracy, the 3.1$\sigma$ confidence of the BC pair is irrelevant. Hence, we label the entire system as undetermined for now, although it is obviously very likely that the whole system is indeed bound.

\textbf{J02545247-0709255}
The primary is an SB2 binary with an 11.8 day period \citep{Torres2002}. Since the angular separation is inside of our inner cut-off, we count the system as a binary here (unconfirmed, since it has not yet been confirmed to share a common proper motion) for the purpose of multiplicity statistics, although it is really a triple system. To estimate the mass of the primary for the purpose of deriving the mass ratio to the tertiary component detected in our images, we adopt the mass ratio of 0.58 derived from the spectroscopic orbit and use the mass corresponding to an M3 spectral type for the primary.

\textbf{J03032132-0805153}
A possible wide companion to J03032132-0805153 at 172\arcsec\ with similar proper motion is noted in WDS \citep{Mason2001}.

\textbf{J03033668-2535329}
One or both of the stars in this system must be highly variable. The magnitude differences between the primary and secondary vary from 5.14 mag to 1.72 mag between different observations in \citet{Bergfors2010} and here. In contrast to the individual note on this system in \citet{Bergfors2010}, we assume here that the designation LTT 1453 refers to the primary star.

\textbf{J03042184+2118154}
The close ($\sim$0.40\arcsec) companion to J03042184+2118154 is counted as an unconfirmed binary companion, since it has not yet been tested for common proper motion.

\textbf{J03050976-3725058}
This target, like J06583980-2021526, gives conflicting information about whether the AB pair is physically bound. The brightness and color are fully consistent with B being an M3.5 companion to the M2.0 primary. Furthermore, the separation is only 0.2\arcsec, and the whole field is otherwise empty. This indicates physical companionship. However, the deviation from the expectation of a background source is only 1.1$\sigma$, and the companion has moved 3.6 times closer to the background expectation than its original position where the motion has a 14$\sigma$ confidence. This, by contrast, suggests a chance alignment. Since the magnitude of motion is not inconsistent with what could be expected for orbital motion of a real physical pair, we keep it as binary in our analysis, but we note that further observations will be necessary in order to resolve this issue satisfactorily.

\textbf{J03283893-1537171}
Also known as GJ 3228A, this star has a wide binary companion at 16\arcsec\ separation \citep[e.g.][]{Weis1991}.

\textbf{J03323578+2843554}
This is a close triple system discovered with AstraLux, where in particular the tertiary component is difficult to fit for. Since the relative positions of the BC pair components are therefore clearly dominated by noise scatter in both separation and position angle, we set the error of each epoch to the scatter among the data points and make no assessment of orbital motion for the BC pair (although by contrast, clear orbital motion is seen of the pair with respect to the A component).

\textbf{J03324345-0855391}
J03324345-0855391 is a component in a system with a rather complex history, as it is a spectroscopic binary with a DA white dwarf with a 0.2 day period and has undergone a common envelope phase in the past \citep[e.g.][]{Davis2010}. Hence, it cannot count as a regular M-star, and is not included in the statistical analysis.

\textbf{J03360868+3118398}
This star has been classified as a possible T Tau star in the Taurus-Auriga region by \citet{Li1998}.

\textbf{J03394784+3328306}
Also known as GJ 9119 B, this star is a wide companion to GJ 9119 A at about 14\arcsec\ as noted in e.g. the WDS catalog \citep{Mason2001}. It is however single in the separation range covered by the AstraLux data.

\textbf{J03415581-5542287}
J03415581-5542287 and J03415608-5542408 are separated by only 12\arcsec\ and have similar estimated spectroscopic distances (14 and 20 pc, respectively) in \citet{Riaz2006}, hence it is likely that they form a physical pair. J03415581-5542287 is a close ($\sim$0.61\arcsec) binary in our data, so the system is likely triple in reality, although common proper motion of all components has yet to be proven.

\textbf{J03423180+1216225}
Since the two epochs of observation were acquired over a baseline of only a few months, there is not yet sufficient motion to confirm common proper motion of the companion that was detected with AstraLux. However, the separation between the components is rather small ($\sim$0.86\arcsec), and the fact that the companion is clearly detected in $z^{\prime}$ but too faint in $i^{\prime}$ implies that it must be very red, as expected for a real companion. Hence, we count it as a binary system for statistical purposes.

\textbf{J03461399+1709176}
This star has a possible wide companion at about 8\arcsec\ as noted in e.g. the Washington Double Star (WDS) catalog \citep{Mason2001}. It is however single in the separation range covered by the AstraLux data.

\textbf{J03472333-0158195}
\citet{Lopez2006} classify this star as a member of subgroup B4 (a nearby young moving group), with an estimated age of $\sim$100 Myr.

\textbf{J03591438+8020019}
A companion to J03591438+8020019 was detected on two separate occasions with AstraLux, in both cases the detection is rather tentative, but since it is detected twice with consistent properties we count it as a genuine detection. However, due to the poor quality of the fit and the fact that the two epochs of observation are only separated by 3 months, we do not try to acquire two epochs of astrometry, but merely quote the astrometry as the mean (and standard deviation) of the two epochs.

\textbf{J04143060+2851298}
J04143060+2851298 and J04143109+2851518, both from the \citet{Riaz2006} sample, are separated by only 23\arcsec\, and although the estimated spectroscopic distances of 55 and 72~pc are not fully equal, they are consistent to within the 37\% error. In addition, they have very similar proper motions \citep[e.g.][]{Roeser2010} and thus probably form a physical pair.

\textbf{J04244260-0647313}
Although this star appears single in our images, it has been identified as a three-component spectroscopic multiple system in \citet{Shkolnik2010}. The largest semi-major axis in the system is constrained as $<$0.25 AU, which at a distance of 35 pc corresponds to approximately $<$7 mas, hence we should indeed not expect to see any of these components in the AstraLux images.

\textbf{J04305203-0849193}
A known component of the wide binary Konigstuhl 2AB with a separation of 20\arcsec\ \citep{Caballero2007}, J04305203-0849193 is single closer in, in the AstraLux images.

\textbf{J04373746-0229282}
Better known as GJ 3305, this binary is part of the $\beta$ Pic moving group \citep{Zuckerman2001}, and is also bound to 51 Eri in a wide orbit \citep{Feigelson2006} with a projected separation of 66\arcsec. The binarity of GJ 3305 was first reported by \citet{Kasper2007}, and has been observed with AstraLux on several occasions.

\textbf{J04374563-0119118}
The X-ray flux of this star is probably due to the fact that it is a symbiotic star, as identified by e.g. \citet{Belczynski2000}, hence we consider it as a contaminant to the \citet{Riaz2006} sample.

\textbf{J04465175-1116476}
J04465175-1116476 has a companion that has not yet been tested for common proper motion. Since the brightness and color match expectations, the system counts as an unconfirmed binary.

\textbf{J05100427-2340407}
Both J05100427-2340407 and J05100488-2340148 are binaries in the AstraLux images, so since they are separated by only 27\arcsec\ on the sky and have very similar proper motions \citep[e.g.][]{Roeser2010}, this is likely a quadruple system in reality. Common proper motion remains to be confirmed for both of the close components.

\textbf{J05130132-7027418}
There is a companion to J05130132-7027418 in the AstraLux images which has a brightness and color that is consistent with a physical companion. It counts as an unconfirmed binary here, since it has not yet been tested for common proper motion.

\textbf{J05195412-0723359}
J05195412-0723359 and J05195513-0723399, both from the \citet{Riaz2006} sample, are separated by only 16\arcsec\ and have comparable estimated distances (59 and 70 pc), as well as quite similar proper motions \citep[e.g.][]{Roeser2010}, and thus likely form a physical pair.

\textbf{J05225705-0850119}
This star has been classified as a probable T Tau star in \citet{Alcala1996}.

\textbf{J05243648-0535175}
There is a possible companion at 7.8\arcsec\ noted in WDS \citep{Mason2001}, but inside the AstraLux field of view, the star appears single.

\textbf{J05301858-5358483}
The AstraLux images resolve this system into a likely triple, where all components have brightnesses and colors consistent with a physically bound system. Since it has only been observed in one epoch, common proper motion has not yet been established.

\textbf{J05320450-0305291}
J05320450-0305291 is identified as a $\beta$ Pic member in \citet{daSilva2009}. It is also known as V1311 Ori.

\textbf{J05323611-0523010}
Also known as HR Ori, this star has been classified as a probable T Tau star in \citet{Alcala1996}.

\textbf{J05343767-0543044}
J05343767-0543044 has been classified as a likely member of the $\sim$1 Myr Orion OBIc/d association \citep{Stassun1999}. The companion that was detected with AstraLux has not yet been confirmed to share a common proper motion, but its brightness and color are consistent with what would be expected for a physical companion, hence it is counted as such in the statistical analysis.

\textbf{J05344858-3239362}
All components of this triple system have colors and brightnesses consistent with expectation for physical companions. Common proper motion has however not yet been demonstrated.

\textbf{J05350429-0508125}
This star has been classified as a probable T Tau star in \citet{Alcala1996}. It is also known as V1321 Ori.

\textbf{J05355975-0616065}
Also known as V1178 Ori, this star has been classified as a probable T Tau star in \citet{Alcala1996}. It has a companion that is as of yet unconfirmed with regards to common proper motion, but since the brightness and color are well consistent with expectations for a real companion, the system is counted as binary in the statistics.

\textbf{J06002304-4401217}
The two components resolved by AstraLux have nearly equal brightnesses and colors. Hence, this counts as an unconfirmed binary, since the system has only been observed in one epoch so far.

\textbf{J06061742-2754050}
This star is likely a binary companion to the nearby K1-type star HD 41842 at 20\arcsec\ separation, as noted in WDS \citep{Mason2001}.

\textbf{J06112997-7213388}
Since the separation between the components of this system is very small ($\sim$0.16\arcsec), it is almost certainly a physical binary, although common proper motion has yet to be demonstrated.

\textbf{J06134171-2815173}
This binary consists of two components with almost equal brightnesses and colors. Although common proper motion has not yet been demonstrated, it thus counts as binary in the statistical analysis.

\textbf{J06134539-2352077}
The close binary pair which was resolved by AstraLux is likely itself a wide companion to the nearby G5-type star HD 43162 at 25\arcsec\ separation, as noted in WDS \citep{Mason2001}. Hence the system is likely triple, although common proper motion has not yet been deomnstrated.

\textbf{J06161032-1320422}
In \citet{Bergfors2010}, we detected a close binary companion to J06161032-1320422 at 190 mas separation. In a re-observation of the target from February 2010, there is no resolved companion detected but only a slightly extended primary PSF, hence the binary separation probably decreased in the meantime.

\textbf{J06234024-7504327}
Although not yet confirmed to share a common proper motion, the companion to J06234024-7504327 is both relatively close $\sim$0.57\arcsec\ and has consistent brightness and color to what should be expected for a physical companion.

\textbf{J06262932-0739540}
The motion in the J06262932-0739540 system is not yet sufficient to determine whether the two components share a common proper motion, but since the separation is small ($\sim$0.47\arcsec) and the colors and brightnesses of the components are consistent with expectation, it is likely that they form a physical pair. The system counts as binary in the statistical analysis.

\textbf{J06281861-0110504}
The companion that was detected in the AstraLux images has not yet been confirmed as a physical companion through common proper motion. However, the two components have nearly equal brightness and colors, and a relatively small separation of $\sim$1.4\arcsec\, hence it is very likely that it is a real physical binary, and it is counted as such here for statistical purposes.

\textbf{J06351837+4745366}
J06351837+4745366 is an unconfirmed binary with regards to common proper motion. Since the brightness and color of the companion is consistent with expectation, we count it as binary in the statistical analysis.

\textbf{J06583980-2021526}
Out of the targets that have been classified as 'undetermined' with respect to common proper motion, the J06583980-2021526AB pair is one out of two that have reasonable (or even large) chances of being chance alignments rather than physical pairs. The fainter candidates J06583980-2021526C and D at larger separations reported in \citet{Bergfors2010} are clearly identified as background contaminants here, given both the colors and proper motions. However, J06583980-2021526B gives conflicting information. On one hand, its brightness and color are fully consistent with it being an almost equal-mass companion to the primary, which would be a very unusual coincidence for a background object, especially at a separation of only 1.4\arcsec. On the other hand, from an astrometric viewpoint, the B component is fully consistent with the expectation for a background object. The deviation from the background expectation is 0.4$\sigma$, and the B component has moved to a position that is 7.4 times closer to the background expectation than its first-epoch position, with a confidence for the motion of 7.5$\sigma$. This is also unlikely to happen by chance, hence the evidence is divergent. Since the magnitude of motion is not inconsistent with what could be expected for orbital motion of a real physical pair, we keep it as binary in our analysis, but we note that further observations will be necessary in order to resolve this issue satisfactorily.

\textbf{J07115917-3510157}
The two components of J07115917-3510157 have nearly equal brightnesses and colors, hence it is very likely that they form a physical pair. They have yet to be demonstrated to share a common proper motion.

\textbf{J07210894+6739590}
In addition to the close binarity discovered in the AstraLux data, there is a star at 21\arcsec\ separation (HIP 35628), but due to its very different proper motion, it is likely physically unrelated to the J07210894+6739590 system \citep{Reid2007}.

\textbf{J07282116+3345127}
Although the star counts as single in our AstraLux data, it is in reality a spectroscopic binary with less than 0.55 AU semi-major axis, as noted by \citet{Shkolnik2010}.

\textbf{J07382951+2400088}
A 2.4\arcsec\ possible companion to J07382951+2400088 is noted in WDS \citep{Mason2001}, but our AstraLux images show no such companion. 

\textbf{J07505369+4428181}
There is a companion to J07505369+4428181 which has a color and brightness that is consistent with what should be expected if it was physcially bound. Common proper motion has not yet been tested for the system.

\textbf{J08082487+4347557}
Although the star counts as single in our AstraLux data, it is in reality a spectroscopic binary with less than 0.02 AU semi-major axis, as noted by \citet{Shkolnik2010}.

\textbf{J08224744-5726530}
As noted in \citet{Bergfors2010}, this is a triple system with component C outside of 6\arcsec. Hence, we do not include component C in Table \ref{t:multep}, but we note that at 8.429$\pm$0.001\arcsec\ separation in 2008.88 and 8.374$\pm$0.003\arcsec\ in 2010.09, there is statistically significant orbital motion also for this component.

\textbf{J08310177+4012115}
For astrometric analysis of J08310177+4012115, we compare our AstraLux measurement with the data from the original discovery as listed in WDS \citep{Mason2001}. Since there are no error bars quoted, it is difficult to formally establish physical companionship. However, even if we assume that the errors are $\pm$0.5\arcsec\ and $\pm$5$^{\rm o}$, which should be very conservative, the background hypothesis can still be firmly excluded with $\sim$5$\sigma$ confidence. We therefore count it as a confirmed companion.

\textbf{J08412528-5736021}
The two components of this system have nearly equal colors and brightnesses, hence they very likely form a physical pair, althouhg they have not yet been tested for common proper motion.  

\textbf{J08472279-4047381} 
This star is a known eclipsing binary with a period of 1.6219 days \citep{Parihar2009}, but has no additional companions in the AstraLux field of view.

\textbf{J08475676-7854532}
More commonly known as EQ Cha, this star appears single in the AstraLux images, but has most likely been partially resolved as a close binary in previous imaging campaigns \citep{Kohler2002,Brandeker2006}. With an angular separation as small as 40 mas in 2003, it is fully plausible that the companion exists but was too close to be resolved by AstraLux in 2010.

\textbf{J08483696-1353087}
J08483696-1353087 and J08483645-1353083 are separated by only 7\arcsec, the former is a close binary (as of yet unconfirmed by common proper motion) and the latter is single, so this is likely a triple system in reality.

\textbf{J09053033-4918382}
This star has a probably brown dwarf companion, but since only one epoch of data exists, this is still unconfirmed.

\textbf{J09121259-2555025}
J09121259-2555025 has the alternative identifier CD-25 6962B in SIMBAD, implicitly implying companionship with the G2-type star CD-25 6962 at 14\arcsec\ separation. However, given the apparently very different proper motions of the components \citep[e.g.][]{Roeser2010}, we consider such a companionship questionable.

\textbf{J09164398-2447428}
We resolve this star into a close binary (as of yet unconfirmed through common proper motion). It was recently classified as a classical Cepheid based on light curve analysis \citep{Christiansen2008}. We assume that this is a mis-classification, given that the unresolved spectral type is M0.5V.

\textbf{J09180165-5452332}
J09180165-5452332 has a close ($\sim$0.49\arcsec) companion to which it is probably physically bound, although this still needs to be confirmed through common proper motion.

\textbf{J09365782-2610111}
The companion discovered with AstraLux is probably physical given the rather small separation ($\sim$0.39\arcsec), but so far only one epoch of imaging exists.

\textbf{J09394631-4104029}
This star is listed as a Tycho double star in the WDS catalog \citep{Mason2000} with a separation of 600 mas and equal brightness. However, it appears entirely single in our AstraLux data. Although we cannot strictly rule out that the companion is presently too close to be resolved with AstraLux (e.g. due to a close to edge-on orbit and unfortunate timing of the observation), we consider it relevant to regard the possible companion as unconfirmed at present.

\textbf{J09423823-6229028}
The two components of this binary have almost equal brightnesses and colors, hence they are very likely to form a physical pair. Common proper motion has not yet been tested.

\textbf{J10023100-2814280}
J10023100-2814280 has a relatively close ($\sim$0.56\arcsec) companion which is probably real, although this has yet to be confirmed with a proper motion test.

\textbf{J09583428-4625300}
While the system appears single in the AstraLux images, it is in reality a spectroscopic binary with a 1.88 day period \citep[e.g.][]{Diaz2007}. Such a small orbit is obviously consistent with a non-detection by AstraLux.

\textbf{J10162867-0520320}
In this previously known system, J10162867-0520320 is a spectroscopic binary of two M-dwarfs components, which is in turn a visual companion at 3.2\arcsec\ separation from a white dwarf (WD). Furthermore, the WD is itself a spectral binary, making the system quadruple, with a demonstrated physical companionship of all four components \citep{Vennes1999}. The 3.2\arcsec\ pairing is easily distinguished in the AstraLux image, but the spectroscopic pairs remain unresolved. Because the system must primordially have had a pre-WD primary much more massive than J10162867-0520320 (given that the main sequence lifetime of an M0 star is longer than the Hubble time), we do not count it as an M-star binary for any of our statistical purposes. The binary is listed as having confirmed proper motion in Table \ref{t:multep}, this is based on the result in \citet{Vennes1999} and not our comparison with the astrometric point in the literature; since no error bars are listed for that point and since we only get rather marginally significant results if we assume reasonable errors of $0.1$\arcsec\ and 1$^{\rm o}$, we do not draw any new conclusions on the basis of such a comparison.

\textbf{J10181387-2028413}
The star is single in the AstraLux images, but has a known wide companion at 32\arcsec\ separation \citep{Deacon2007}.

\textbf{J10423011-3340162}
The source at 2.4\arcsec\ separation, first reported in \citet{Neuhauser2000}, is an already known background star \citep{Lowrance2005}, a conclusion which we can confirm at $\sim$28$\sigma$ confidence.

\textbf{J11091380-3001398}
This previously known binary \citep[e.g.][]{Webb1999} is part of the TW Hya association, and is also known as TWA 2.

\textbf{J11102788-3731520}
This previously known binary \citep[e.g.][]{Brandeker2003} is part of the TW Hya association, and is also known as TWA 3.

\textbf{J11240434+3808108}
There is a possible brown dwarf companion to J11240434+3808108 at 8\arcsec\ separation noted by \citet{Reid2007}, but it is single within the AstraLux field of view.

\textbf{J11254754-4410267}
J11254754-4410267 has a relatively close ($\sim$0.55\arcsec) companion that is likely physically bound, but so far there is only a single epoch of data available.

\textbf{J11315526-3436272}
This system has been identified as a member of the TW Hya association \citep{Kastner1997} and is known as TWA 5. It contains a previously known brown dwarf TWA 5B, first suggested by \citet{Lowrance1999} and later confirmed by e.g. \citet{Brandeker2003}. The latter also reported the presence of a close companion TWA 5Ab at 54 mas from the primary, independently detected by \citet{Macintosh2001}. We detect TWA 5B in our AstraLux data, but not TWA 5Ab, which implies that the projected separation was probably smaller in 2010 than it was in 2000. The orbit of TWA 5Aa/Ab has been analyzed in detail by \citet{Konopacky2007}. Using the orbital parameters that they determine, we deduce that at epoch 2010.11, the projected separation should be only about 18 mas, which is well consistent with our non-detection of TWA 5Ab. We use the best-fit total mass of 0.71 $M_{\sun}$ for TWA 5Aa/Ab from \citet{Konopacky2007}.

\textbf{J12045611+1728119}
Common proper motion is shared between J12045611+1728119 and HIP 58919 at 23\arcsec\ separation, hence they constitute a wide physical pair \citep{Lepine2007}.

\textbf{J12062214-1314559}
Due to the small separation ($\sim$0.42\arcsec) of the companion detected in the AstraLux images, it is likely a physical companion, although common proper motion has not yet been tested.

\textbf{J1206557+700749}
Aside from the spatially resolved companion, the primary in the J1206557+700749 system is a known spectroscopic binary with a semi-major axis less than 0.03~AU. This is well consistent with our AstraLux data, where the primary is significantly brighter than the secondary, despite having about equal spectral type. Hence, the system is a triple in reality.

\textbf{J12134173-1122405}
A companion to J12134173-1122405 is detected in the AstraLux images which has not yet been tested for common proper motion, but the color and brightness is consistent with the expectation for a physically bound companion.

\textbf{J12173945-6409418}
The two targets J12173945-6409418 and J12174012-6409389 are physically bound, as shown by common proper motion analysis. However, this means that one of the targets likely has an error in the \citet{Riaz2006} spectral type determination, since J12174012-6409389 is classified as M2 but is fainter than J12174012-6409389, which is classified as M3.5. The fact that J12174012-6409389 is fainter holds true both in 2MASS and in the AstraLux images. Both components are resolved as close binaries with AstraLux, so we consider it a strong candidate for a quadruple system, with J12173945-6409418A as the primary. Concerning which spectral type determination is incorrect, we consider that it is more likely that J12174012-6409389 is mis-classified, as the flux ratio to the close companion is closer to unity than in the J12174012-6409389 case. We thus set the spectral type of J12174012-6409389A to M3$\pm$1, and determine the other spectral types on the basis of flux ratios. Component Bb is not visible in the $i^\prime$ data, but becomes visible in $z^\prime$ thanks to the higher Strehl ratio and intrinsic brightness of the component. A fifth object is visible in the field, but its color reveals it to be a likely background star.

\textbf{J12345629-4538075}
J12345629-4538075 is better known as TWA 16. The presence of a close companion to this star was noted in \citet{Zuckerman2001} which seems to be consistent with the companion that we detect. However, since no explicit astrometric information is given in \citet{Zuckerman2001} other than a rough estimation of the separation (0.67\arcsec), it is not possible to test whether the companion shares a common proper motion with the primary at this point. The brightness and color are consistent with a physical companion, hence the system counts as binary in the statistics.

\textbf{J12351726+1318054}
The position angle for this system given in \citet{Law2006} is actually 257.0$^{\rm o}$, but since the components are fairly similar in brightness and there is a fake triplet effect, we assume that there can be a 180$^{\rm o}$ phase shift present in the \citet{Law2006} data (or alternatively in both of our epochs). The angular motion in our two AstraLux images is 1.8$^{\rm o}$ in about 0.5 years. If taken at face value, the \citet{Law2006} data point would imply 198.4$^{\rm o}$ angular motion in 4.5 years, corresponding to 22$^{\rm o}$ per half-year, which is an order of magnitude too large. On the other hand, if we subtract 180$^{\rm o}$ from the \citet{Law2006} position angle to get 77.0$^{\rm o}$, then the angular motion is 2.0$^{\rm o}$ per half-year, which is perfectly consistent with our measurements. Hence, we adopt the latter value for our analysis.

\textbf{J12392104-5337579}
The two components of J12392104-5337579 that were detected in the AstraLux images have almost equal brightnesses and color, hence they are very likely physically bound. Common proper motion has not yet been tested.

\textbf{J12485345+1204326}
J12485345+1204326 is a wide (76\arcsec) common proper motion companion to HIP 62536 \citep{Lepine2007}, but is single in the AstraLux images.

\textbf{J12533626+2247354}
The star shows photometric variability with a 1.9 day periodicity \citep{Norton2007}.

\textbf{J12565215+2329501}
J12565215+2329501 and J12565272+2329506 constitute a known 8\arcsec\ binary pair in the \citet{Reid2007} sample.

\textbf{J13013268+6337496}
Aside from the binary companion detected in the AstraLux images, the star has a probable wide companion at 119\arcsec, as noted in WDS \citep{Mason2001}.

\textbf{J13015919+4241160}
The companion to the J13015919+4241160 is clearly inconsistent with a background contaminant, but the astrometry is also inconsistent with a simple orbital motion of B around A -- for instance, the separation decreases from June 2008 to February 2009, but then increases again to June 2009. This could for instance imply that either A or B is a close unresolved binary, where the photo-center shifts on a shorter timescale than that of the AB orbit.

\textbf{J13022691-5200507}
Due to the very compact arrangement of the three components of J13022691-5200507, it is highly probable that this is a physically bound triple system. Futhermore, the system is conspicuously close to the close binary J13025257-5201384 both on the sky and in distance (52 versus 59~pc), hence this is a candidate quintuple system. Only one epoch of imaging exists, so common proper motion has not yet been demonstrated.

\textbf{J13082484+3019094}
Aside from the binary companion detected in the AstraLux images, the star has a possible wide companion at 101\arcsec, as noted in WDS \citep{Mason2001}.

\textbf{J13093495+2859065}
Also known as GJ 1167 A, this star has a wide companion at about 190\arcsec\ according to WDS \citep{Mason2001}.

\textbf{J13120689+3213179}
J13120525+3213332 and J13120689+3213179 constitute a known 26\arcsec\ binary pair in the \citet{Reid2007} sample, and in addition, each of the components are discovered as close binaries in the AstraLux data. Hence, the system is a very likely quadruple system, although the close pair of J13120689+3213179 has yet to be formally confirmed as physically bound.

\textbf{J13151846-0249516}
Due to the small separation of the companion to J13151846-0249516 it is very likely a physical binary, but common proper motion has not yet been tested for.

\textbf{J13195689-6831142}
J13195689-6831142 has a companion detected in the AstraLux images. Since the companion is relatively close ($\sim$0.88\arcsec) and has a brightness and color that is consistent with expectation, it is probably a physical pair. Only one epoch of images exists so far.

\textbf{J13293209+5142114}
The B component seen in the AstraLux images appears to be a very close binary itself due to a PSF extension visible in five separate epochs. However, we count the system as a regular binary here since we do not get a converging fit for the closer pair.

\textbf{J13313493+5857171}
We count J13313493+5857171 as single close in, although its PSF appears somewhat extended in the 2009.13 epoch. However, it is a likely companion to the nearby G2-star HD~117845 at 11\arcsec\ separation. HD~117845 has been included in the AstraLux field and appears to be itself a close binary with a $\sim$0.5\arcsec\ separation.

\textbf{J13345147+3746195}
Although the close binarity of J13345147+3746195 was reported in \citet{Daemgen2007} and our AstraLux data have a similar sensitivity, the star appears single in our images. Since the separation was only 82 mas in 2006, it has presumably moved inward since then.

\textbf{J13493313-6818291}
Due to the compact configuration of the three components resolved by AstraLux, the probability is very high for this to constitute a physical triple system. Common proper motion has not yet been tested for.

\textbf{J13534589+5210298}
The secondary detected in the AstraLux images is perhaps itself a close binary, as it appears extended in epochs 2008.64 and 2009.42. However, we count it as a single component of the system here.

\textbf{J14134677+4618227}
J14134677+4618227 is possibly a wide companion to HIP 69518 at 83\arcsec\ \citep{Reid2007}. It is however single in the AstraLux field of view.

\textbf{J14201961+2758563}
This star is variable, and has been classified as a classical Cepheid based on this variability \citep{Akerlof2000}. We assume that this is a mis-classification, given that the spectral type is K5.

\textbf{J14204953+6049348}
Although the star is single in our AstraLux images, it is known to be a spectroscopic binary with a semi-major axis of 0.01 AU \citep{Shkolnik2010}.

\textbf{J14402293+1339230}
This late-type (M7.5--M8.0) object is classified as a brown dwarf in SIMBAD. It is however unclear what this classification is based on, as our comprehensive search through the full published body of literature available on the object did not turn up any references or justification to such a classification.

\textbf{J14433804-0414354}
The two components resolved by AstraLux have very similar brightnesses and colors, hence physical companionship is probable, although common proper motion has not yet been tested for.

\textbf{J15090696+5904282}
J15090696+5904282 and J15090808+5904258 are separated by only 9\arcsec\ and have similar estimated spectroscopic distances (30 and 39 pc, respectively) in \citet{Riaz2006}, hence it is likely that they form a physical pair.

\textbf{J15235385+5609320}
A probable wide companion to this star at approximately 68\arcsec\ separation is noted in WDS \citep{Mason2001}.

\textbf{J15370409+3748275}
The star displays photometric variability at a 1.2 day period \citep{Norton2007}.

\textbf{J15530484+4457446}
The close binary that we discovered in the AstraLux images was not seen in the \citet{Daemgen2007} study with a similar spatial resolution, which implies that the projected separation of the system probably increased during the intermediate period. 

\textbf{J15553178+3512028}
In addition to the astrometric points listed in Table \ref{t:multep}, there is another data point from 2005 in \citet{Law2008}. This point is essentially at equal epoch to the \citet{Daemgen2007} point, but the position angle differs by 88$^{\rm o}$ in the two cases. While the \citet{Daemgen2007} data point is fully consistent with all other astrometric points of the target, the \citet{Law2008} point is fully inconsistent in this regard. The close to 90$^{\rm o}$ offset could imply some trivial trigonometric error in \citet{Law2008}. Here we simply exclude the point from our analysis.

\textbf{J16291031+7804399}
Like J10162867-0520320, this is a previously known system with a white dwarf component \citep{Farihi2010}. Hence, we do not include this system in the binary fraction statistics.

\textbf{J16494292+2220037}
The star shows photometric variability with a 22.9 day periodicity, which might imply that it is an eclipsing binary \citep{Norton2007}.

\textbf{J16552880-0820103}
Also known as GJ 644, this is a well-known and much studied triple system \citep[e.g.][]{Segransan2000}, of which we resolve the wider AB pair (the secondary is a spectroscopic binary with a 3 day period). In addition, there are two much wider components known to share a common proper motion with GJ 644, making the system a likely quintuple. Since there are so many existing astrometric points of the AB pair in the literature, we do not give an extensive list in Table \ref{t:multep}, but rather refer the reader to \citep{Segransan2000}, where there is a full orbital analysis of the close triple system.

\textbf{J16590962+2058160}
The star displays photometric variability at a 4.1 day period \citep{Norton2007}.

\textbf{J17111769+1245408}
J17111769+1245408 and J17111841+1245080, both from the \citet{Riaz2006} sample, are separated by only 34\arcsec\ and have comparable estimated distances (69 and 82 pc), as well as similar proper motions \citep[e.g.][]{Roeser2010}, and thus likely form a physical pair.

\textbf{J17292722+3524048}
This star is variable, and has been classified as a classical Cepheid based on this variability \citep{Akerlof2000}. We assume that this is a mis-classification, given that the spectral type is K5.

\textbf{J17380077+3329457}
The B component of the system appears to be a close binary itself in several different epochs of AstraLux imaging, but the suspected Ba/Bb pair is not sufficiently resolved to get a converging binary fit, hence we treat the AB system as a regular binary in this study.

\textbf{J18153459+1614253}
A possible wide companion to J18153459+1614253 at 13\arcsec\ is noted in WDS \citep{Mason2001}. The star is however single within the AstraLux field of view.

\textbf{J18351833+4544379}
Also known as GJ 720 A, this star has a wide companion with a demonstrated common proper motion at about 112\arcsec\ according to \citet{Lepine2007}.

\textbf{J18355276+1659057}
J18355276+1659057 is possibly a wide companion to HIP 91159 at 35\arcsec\ separation, as noted in \citet{Reid2007}.

\textbf{J18464053-0916238}
The primary star has two candidate companions in the field of view. The fainter candidate C is most likely a background star, since its color is entirely inconsistent with a late M-type star. The brighter candidate B is more uncertain. It's colors are fully consistent with the brightness contrast to the primary for a physical companion, implying a spectral type of M5. It is also visible in 2MASS \citep{Skrutskie2006}, which enables to perform a rough CPM test. The test indicates that B is inconsistent with a static background object, again implying physical companionship. However, the motion is also poorly consistent with orbital motion, with a sky-projected speed of order 2~AU per year -- highly unlikely for such a wide and low-mass binary. This implies either under-estimated uncertainties in the data, or an unusual astronomical coincidence. With the small separation and relatively high brightness difference of the candidate companion, it is just at the detection limit of 2MASS, and is only reliably detectable in K-band. Perhaps there is a systematic noise component beyond the quoted astrometric 2MASS error in this circumstance. If all the data are taken at face value, another possible interpretation is an unfortunate chance alignment with a local M-dwarf, which has a signficant proper motion of its own. In any case, due to all these uncertainties, we tentatively treat the pair as a candidate binary here, but emphasize that more data is needed to test physical companionship.

\textbf{J18564143+5014071}
J18564143+5014071 and J18564286+5013483 are separated by only 23\arcsec\ and have similar estimated spectroscopic distances (79 and 77 pc, respectively) in \citet{Riaz2006}, as well as very similar proper motions \citep[e.g.][]{Roeser2010}, hence it is likely that they form a physical pair.

\textbf{J19105480+3017476}
The AstraLux images reveal three components in the J19105480+3017476 system. Due to their very compact configuration, the triple system is very likely physically bound. Only one epoch of imaging exists so far.

\textbf{J19213210+4230520}
The very close companion resolved by AstraLux is probably physically bound, but this has yet to be tested through proper motion analysis.

\textbf{J19425324-4406278}
A close companion to J19425324-4406278 is detected which is counted as a physical companion for statistical purposes, although this has not yet been demonstrated through common proper motion.

\textbf{J19432464-3722108}
Only one epoch of imaging exists for this binary system, hence common proper motion has not yet been confirmed. It is counted as an unconfirmed binary here.

\textbf{J20003177+5921289}
Aside from the close companion detected in the AstraLux data, J20003177+5921289 possibly has a wide companion at 14\arcsec, as noted in WDS \citep{Mason2001}.

\textbf{J20100002-2801410}
Given the relatively small separation ($\sim$0.62\arcsec) and the similar brightnesses of the two components in the J20100002-2801410 system, it is likely that they form a physical pair. However, only one epoch of imaging exists at this point.

\textbf{J20163382-0711456}
Due to the very small separation of the detected companion to J20163382-0711456, this is likely to be a physical binary, although a second epoch to confirm common proper motion has not yet been acquired.

\textbf{J20500010-1154092}
J20500010-1154092 has a close companion detected with AstraLux, but it has not yet been confirmed through common proper motion. However, since the separation is quite small ($\sim$0.47\arcsec) and the brightness and color of the companion is consistent with expectations, the system counts as binary for statistical purposes.

\textbf{J20564846-0450490}
This star has a wide companion at about 14\arcsec\ as noted in e.g. the WDS catalog \citep{Mason2001}. It is however single in the separation range covered by the AstraLux data.

\textbf{J20581756+1541315}
J20581756+1541315 and J20581836+1541211, both from the \citet{Riaz2006} sample, are separated by only 10\arcsec\ and have comparable estimated distances (144 and 119 pc), as well as similar proper motions \citep[e.g.][]{Roeser2010} and thus likely form a physical pair.

\textbf{J21091375-0814041}
The binary J21091375-0814041 is as of yet uncomfirmed with regards to common proper motion, but the components have similar colors and brightnesses, and a relatively small separation of $\sim$0.97\arcsec\, hence the pair is probably physically bound, and counts as such in the statistical analysis.

\textbf{J21154192+1746242}
This star possibly has a wide companion at about 9\arcsec\ as noted in e.g. the WDS catalog \citep{Mason2001}. It is however single in the separation range covered by the AstraLux data.

\textbf{J21205172-0301545}
J21205172-0301545 has a companion with brightness and color that is consistent with expectation, but no common proper motion test has yet been done.

\textbf{J21295166-0220070}
The close companion to J21295166-0220070 has consistent brightness and color with the expectation for a physical companion. Only one epoch of images exists, hence common proper motion has yet to be confirmed.

\textbf{J21365560-0840313}
Due to the small separation ($\sim$0.24\arcsec) of the companion detected in the AstraLux images, physical companionship is very likely, although only one epoch of images exists so far.

\textbf{J22014336-0925139}
The two components of the J22014336-0925139 system that have been resolved with AstraLux have very similar brightnesses and colors, and it is therefore likely that they form a physical binary pair. Common proper motion has not yet been demonstrated.

\textbf{J22171899-0848122}
This system counts as single in our statistics, even though it appears to be a triple system in reality. The reason for this is that the A component was the AstraLux target, and the BC pair at $\sim$7.8\arcsec\ is outside of the completeness cut-off at 6\arcsec. Since the BC pair is visible in the AstraLux images, we can nonetheless analyze it astrometrically. This pair was first reported by \citet{Beuzit2004} with $\rho = 0.978$\arcsec\ and $\theta = 305.8^{\rm o}$. There are no error bars quoted, but if we assume that the quoted precision in decimal places corresponds to the measurement precision, and adopt errors of 5 mas and 0.5$^{\rm o}$, which should be conservative in that circumstance, we find that common proper motion can be confirmed at a 76$\sigma$ confidence level. Hence, physical companionship can be confidently inferred, even if the errors should be substantially larger than what we have assumed.

\textbf{J22232904+3227334}
Also known as GJ 856, this is a well studied binary with several astrometric data points over a long baseline. In Table \ref{t:multep}, we present the AstraLux data as well as one example data point from the literature, and otherwise refer to \citet{Seymour2002} for more information on the astrometric data and a preliminary orbit fit. \citet{Lopez2006} classify the system as a member of subgroup B4, with an estimated age of $\sim$100 Myr.

\textbf{J22545501+2414451}
This star has a probable wide companion at about 73\arcsec\ as noted in e.g. the WDS catalog \citep{Mason2001}. It is however single in the separation range covered by the AstraLux data.

\textbf{J23062378+1236269}
In addition to the previously known binary companion seen in the AstraLux images, J23062378+1236269 is also a spectroscopic binary with a semi-major axis smaller than 0.31 AU \citep{Shkolnik2010}, and has the additional wide companion J23062530+1236570 (also in the sample, a single star in the AstraLux images) at 37.6\arcsec\ with confirmed proper motion \citep{Lepine2007}, hence the system is likely quadruple in reality.

\textbf{J23062928-0502285}
The M7.5 spectral type classification of J23062928-0502285 originates from a spectroscopic analysis in \citet{Gizis2000}. There is some ambiguity to this classification in that \citet{Bouy2003} find it to be a brown dwarf of type L4. However, the latter estimate is based on a single color, so we consider it less reliable.

\textbf{J23101857+1447203}
The star is classified as T Tau in \citet{Li2000}.

\textbf{J23120603+2655579}
As noted in \citet{Lepine2007}, J23120603+2655579 forms a probable wide (14\arcsec) pair with HIP 114543, which is itself a 1\arcsec\ binary. Hence, with the additional component seen in the AstraLux images, the system is likely quadruple.

\textbf{J23205766-0147373}
Although the close binarity of J23205766-0147373 was reported in \citet{Daemgen2007} and our AstraLux data have a similar sensitivity, the star appears single in our images. Since the separation was only 99 mas in 2005, it has presumably moved inward since then.

\textbf{J23230117-0635436}
There is a spectral classification ambiguity for this target, with a K5 classification in \citet{Riaz2006} and a G9III classification in SIMBAD. Regardless of which one is trusted, the star is too early-type to be included in the statistical studies anyway.

\textbf{J23261707+2752034}
A very close ($\sim$0.15\arcsec) companion is detected, which is most likely physically bound, although this has yet to be confirmed with common proper motion.

\textbf{J23314492-0244395}
There is a possible companion at 19\arcsec\ noted in \citet{Lowrance2005}, but it has not been checked for common proper motion.

\textbf{J23342274+2739556}
This star has a possible wide companion at about 63\arcsec\ as noted in e.g. the WDS catalog \citep{Mason2001}. It is however single in the separation range covered by the AstraLux data.

\textbf{J23385413-1246184}
This late-type (M6.5) object is classified as a brown dwarf in SIMBAD. It is however unclear what this classification is based on, as our comprehensive search through the full published body of literature available on the object did not turn up any references or justification to such a classification.

\textbf{J23450477+1458573}
Due to the compact configuration of the three components resolved by AstraLux, this is almost certainly a physically bound triple system. Only one epoch of images exist so far, hence common proper motion has not yet been demonstrated.

\begin{longtable}{lllllllll}
\caption{Properties of all stars observed in the AstraLux campaign.}\\
\hline\hline
2MASS ID & Other name & SpT & D [pc]$^{\rm a}$ & J mag & Mult.$^{\rm b}$ & X-ray$^{\rm c}$ & Sample$^{\rm d}$ & Norte/Sur$^{\rm e}$\\
\hline
\endfirsthead
\caption{continued.}\\
\hline\hline
2MASS ID & Other name & SpT & D [pc]$^{\rm a}$ & J mag & Mult.$^{\rm b}$ & X-ray$^{\rm c}$ & Sample$^{\rm d}$ & Norte/Sur$^{\rm e}$\\
\hline
\endhead
\hline
\endfoot
\hline
\multicolumn{9}{l}{\footnotesize{$^a$Distance is spectroscopic distance except for cases indicated by an acronym. Parallax distances: J52 \citep{Jenkins1952},}}\\
\multicolumn{9}{l}{\footnotesize{H80 \citep{Harrington1980}, A95 \citep{vanAltena1995}, P97 \citep[\textit{Hipparcos}; ][]{Perryman1997}, R02 \citep{Reid2002},}}\\
\multicolumn{9}{l}{\footnotesize{G04 \citep{Gould2004}, R04 \citep{Reid2004}, H06 \citep{Henry2006}, L09 \citep{Lepine2009}, R10 \citep{Reidel2010},}}\\
\multicolumn{9}{l}{\footnotesize{S10 \citep{Smart2010}. Kinematic distances: M05 \citep{Mamajek2005}. The uncertainty in the spectroscopic distances is 37\%.}}\\
\multicolumn{9}{l}{\footnotesize{$^b$Multiplicity status as observed with AstraLux. M: Multiple system. S: Single star. BG: Confirmed or suspected background star.}}\\
\multicolumn{9}{l}{\footnotesize{U: Unclear case.}}\\
\multicolumn{9}{l}{\footnotesize{$^c$Flag for whether X-ray emission is detected at $>$3.3$\sigma$ (Y) or not (N), see Sect. \ref{s:sample} for details.}}\\
\multicolumn{9}{l}{\footnotesize{$^d$This column lists membership of IS or CS (members of neither are marked with N). See Sect. \ref{s:sample} for details.}}\\
\multicolumn{9}{l}{\footnotesize{$^e$Observed with AstraLux Norte (N) or Sur (S).}}\\
\hline
\endlastfoot
J00053484-0607070	&	LP 644-34	&	M2.5 	&	25	&	9.28 	&	S	&	N	&	IS	&	N	\\
J00063925-0705354	&		&	M3.5 	&	14	&	9.86 	&	M	&	N	&	IS	&	N	\\
J00080642+4757025	&		&	M4.0 	&	12	&	8.55 	&	S	&	Y	&	CS	&	N	\\
J00142956+1331086 	&		&	K7.0	&	70 	&	9.34 	&	S	&	Y	&	IS	&	N	\\
J00150240-7250326	&		&	M1.0	&	41	&	8.62	&	M	&	Y	&	CS	&	S	\\
J00155808-1636578	&		&	M4.0	&	9	&	8.74	&	S	&	Y	&	CS	&	S	\\
J00165001-0710157 	&		&	M0.0	&	64 	&	9.58 	&	S	&	Y	&	IS	&	N	\\
J00171443-7032021	&		&	M0.5	&	48	&	9.00	&	S	&	Y	&	CS	&	S	\\
\end{longtable}

\begin{longtable}{lllllll}
\caption{Relative photometry and astrometry of confirmed and unconfirmed physical pairs.}\\
\hline\hline
2MASS ID & $\Delta z^{\prime}$ (mag) & $\Delta i^{\prime}$ (mag) & $\rho$ (\arcsec) & $\theta$ ($^{\rm o}$) & Epoch & New \\
\hline
\endfirsthead
\caption{continued.}\\
\hline\hline
2MASS ID & $\Delta z^{\prime}$ (mag) & $\Delta i^{\prime}$ (mag) & $\rho$ (\arcsec) & $\theta$ ($^{\rm o}$) & Epoch & New \\
\hline
\endhead
\hline
\endfoot
\hline
\hline
\endlastfoot
J00063925-0705354AB	&	1.17$\pm$0.05	&	1.51$\pm$0.01	&	0.250$\pm$0.001	&	5.5$\pm$0.4	&	2008.88	&	Y	\\
J00150240-7250326AB	&	2.65$\pm$0.03	&	2.72$\pm$0.08	&	0.290$\pm$0.009	&	69.1$\pm$0.3	&	2008.86	&	Y	\\
J00250428-3646176AB	&	3.34$\pm$0.23	&	3.65$\pm$0.21	&	0.605$\pm$0.012	&	242.4$\pm$0.3	&	2008.86	&	Y	\\
J00325313-0434068AB	&	2.53$\pm$0.06	&	3.17$\pm$0.02	&	0.428$\pm$0.006	&	179.2$\pm$0.9	&	2008.88	&	Y	\\
J00414141+4410530AB	&	0.36$\pm$0.11	&	0.24$\pm$0.07	&	0.579$\pm$0.001	&	11.4$\pm$0.3	&	2008.59	&	Y	\\
J00485822+4435091AB	&	0.20$\pm$0.11	&	0.17$\pm$0.01	&	1.053$\pm$0.002	&	254.2$\pm$0.3	&	2008.59	&	N	\\
J00503319+2449009AB	&	0.82$\pm$0.04	&	0.90$\pm$0.01	&	1.320$\pm$0.001	&	319.0$\pm$0.3	&	2008.59	&	Y	\\
J01034210+4051158AB	&	0.68$\pm$0.01	&	0.81$\pm$0.01	&	2.477$\pm$0.001	&	96.8$\pm$0.3	&	2008.64	&	N	\\
\end{longtable}

\begin{longtable}{llllllll}
\caption{Derived parameters for confirmed and unconfirmed physical pairs.}\\
\hline\hline
2MASS ID & SpT$_{\rm A}$ & SpT$_{\rm B}$ & $m_{\rm A}$ & $m_{\rm B}$ & Sep. & Per.$^{\rm a}$ & Sample$^{\rm b}$ \\
 &  &  & [$M_{\rm sun}$] & [$M_{\rm sun}$] & [AU] & [yr] & \\
\hline
\endfirsthead
\caption{continued.}\\
\hline\hline
2MASS ID & SpT$_{\rm A}$ & SpT$_{\rm B}$ & $m_{\rm A}$ & $m_{\rm B}$ & Sep. & Per.$^{\rm a}$ & Sample$^{\rm b}$ \\
 &  &  & [$M_{\rm sun}$] & [$M_{\rm sun}$] & [AU] & [yr] & \\
\hline
\endhead
\hline
\endfoot
\hline
\multicolumn{8}{l}{\footnotesize{$^a$Estimated period based on the system mass and approximate semimajor axis. This should only be interpreted on an}}\\
\multicolumn{8}{l}{\footnotesize{order-of-magnitude basis.}}\\
\multicolumn{8}{l}{\footnotesize{$^b$IS denotes Inclusive Sample and CS denotes Constrained Sample, see Sect. \ref{s:sample}. N means the star is in neither sample.}}\\
\hline
\endlastfoot
J00063925-0705354AB	&	M3.0	&	M4.5	&	0.290	&	0.175	&	3.5$\pm$1.3	&	14	&	IS	\\
J00150240-7250326AB	&	M1.0	&	M4.5	&	0.540	&	0.175	&	11.7$\pm$4.4	&	63	&	CS	\\
J00250428-3646176AB	&	M2.5	&	M6.0	&	0.355	&	0.120	&	17.4$\pm$6.5	&	144	&	CS	\\
J00325313-0434068AB	&	M3.5	&	M6.0	&	0.245	&	0.120	&	4.9$\pm$1.8	&	25	&	IS	\\
J00414141+4410530AB	&	M2.0	&	M2.5	&	0.420	&	0.355	&	45.5$\pm$16.9	&	493	&	IS	\\
J00485822+4435091AB	&	M3.0	&	M3.0	&	0.290	&	0.290	&	47.0$\pm$17.5	&	598	&	IS	\\
J00503319+2449009AB	&	M3.5	&	M4.5	&	0.245	&	0.175	&	15.8$\pm$5.9	&	137	&	CS	\\
J01034210+4051158AB	&	K7.0	&	M0.0	&	0.630	&	0.590	&	79.3$\pm$2.0	&	904	&	IS	\\
\end{longtable}

\begin{longtable}{llllllllll}
\caption{Pairs observed over several epochs.}\\
\hline\hline
2MASS ID & $\rho$ (\arcsec) & $\sigma_{\rho}$ & $\theta$ ($^{\rm o}$) & $\sigma_{\theta}$ & Epoch & Ref. & PM$^{\rm a}$ & CPM$^{\rm b}$ & OM$^{\rm c}$ \\
 &  &  &  &  &  &  & (mas/yr) &  &  \\
\hline
\endfirsthead
\caption{continued.}\\
\hline\hline
2MASS ID & $\rho$ (\arcsec) & $\sigma_{\rho}$ & $\theta$ ($^{\rm o}$) & $\sigma_{\theta}$ & Epoch & Ref. & PM$^{\rm a}$ & CPM$^{\rm b}$ & OM$^{\rm c}$ \\
 &  &  &  &  &  &  & (mas/yr) &  &  \\
\hline
\endhead
\hline
\endfoot
\hline
\multicolumn{10}{l}{\footnotesize{$^a$Total proper motion per year.}}\\
\multicolumn{10}{l}{\footnotesize{$^b$Flag for common proper motion (yes or no). `U' refers to unclear cases.}}\\
\multicolumn{10}{l}{\footnotesize{$^c$Flag for significant orbital motion (yes or no). `U' refers to unclear cases.}}\\
\multicolumn{10}{l}{\footnotesize{$^d$A large number of observations exist in the literature; one example listed here (see individual notes).}}\\
\multicolumn{10}{l}{\footnotesize{$^e$See individual note for J12351726+1318054.}}\\
\multicolumn{10}{l}{\footnotesize{$^f$Not included in binary statistics; see individual note.}}\\
\hline
\endlastfoot
J00063925-0705354	&	0.230	&	0.006	&	6.5	&	0.5	&	2008.63	&	This paper	&	143	&	Y	&	Y	\\
	&	0.250	&	0.001	&	5.5	&	0.4	&	2008.88	&	This paper	&		&		&		\\
\\																			
J00325313-0434068	&	0.422	&	0.012	&	180.0	&	2.2	&	2008.63	&	This paper	&	168	&	Y	&	N	\\
	&	0.428	&	0.006	&	179.2	&	0.9	&	2008.88	&	This paper	&		&		&		\\
\\																			
J00414141+4410530	&	0.580	&	0.002	&	10.7	&	0.3	&	2008.03	&	This paper	&	50	&	Y	&	Y	\\
	&	0.579	&	0.001	&	11.4	&	0.3	&	2008.59	&	This paper	&		&		&		\\
\end{longtable}

\section{Background stars}
In Table \ref{t:background}, we summarize the objects detected in the AstraLux images that have been identified as probable (or in some cases near-certain) background objects.

\begin{longtable}{lllllll}
\caption{Properties of probable background stars.}\\
\hline\hline
2MASS ID & $\Delta z^{\prime}$ (mag) & $\Delta i^{\prime}$ (mag) & $\rho$ (\arcsec) & $\theta$ ($^{\rm o}$) & Epoch & Comment \\
\hline
\endfirsthead
\caption{continued.}\\
\hline\hline
2MASS ID & $\Delta z^{\prime}$ & $\Delta i^{\prime}$ & $\rho$ (\arcsec) & $\theta$ ($^{\rm o}$) & Epoch & Comment \\
\hline
\endhead
\hline
\endfoot
\hline
\hline
\endlastfoot
J01365516-0647379AB	&	5.45$\pm$0.01	&	5.11$\pm$0.01	&	5.580$\pm$0.007	&	180.6$\pm$0.3	&	2008.63	&	Off color	\\
J04133532+1541016AB	&	-0.85$\pm$0.43	&	-1.26$\pm$0.27	&	5.609$\pm$0.003	&	289.1$\pm$0.3	&	2008.63	&	Non-CPM	\\
J06141327-0035516AB	&	5.49$\pm$0.29	&	5.18$\pm$0.21	&	5.749$\pm$0.002	&	320.1$\pm$0.3	&	2008.02	&	Off color	\\
J06315741-0832494AB	&	2.51$\pm$0.01	&	2.52$\pm$0.01	&	5.628$\pm$0.003	&	107.7$\pm$0.3	&	2008.02	&	Off color	\\
J06583980-2021526AD	&	6.91$\pm$0.03	&	...	&	5.149$\pm$0.001	&	253.9$\pm$0.3	&	2008.87	&	Unclear case	\\
J07174710-2558554AB	&	6.35$\pm$0.04	&	...	&	5.332$\pm$0.002	&	126.8$\pm$0.3	&	2008.88	&	Unclear case	\\
J07184418-2616172AB	&	8.27$\pm$0.11	&	8.63$\pm$0.14	&	2.733$\pm$0.015	&	247.4$\pm$0.3	&	2010.08	&	Unclear case	\\
J08471906-5717547AB	&	6.78$\pm$0.09	&	7.63$\pm$0.14	&	5.724$\pm$0.014	&	326.6$\pm$0.3	&	2010.09	&	Unclear case	\\
\end{longtable}

\clearpage


\begin{thebibliography}{99}
\bibitem[Akerlof et al.(2000)]{Akerlof2000} Akerlof, C. et al.\ 2000, AJ 119, 1901
\bibitem[Alcala et al.(1996)]{Alcala1996} Alcala, J.M. et al.\ 1996, A\&AS 119, 7
\bibitem[Bate(2009)]{Bate2009} Bate, M.R.\ 2009, MNRAS 392, 590
\bibitem[Bate(2012)]{Bate2012} Bate, M.R.\ 2012, MNRAS 419, 3115
\bibitem[Basu \& Vorobyov(2012)]{Basu2012} Basu, S. \& Vorobyov, E.I.\ 2012, ApJ 750, 30
\bibitem[Belczynski et al.(2000)]{Belczynski2000} Belczynski, K., Mikolajewska, J., Munari, U., Ivison, R.J., \& Friedjung, M.\ 2000, A\&AS 146, 407
\bibitem[Bergfors et al.(2010)]{Bergfors2010} Bergfors, C. et al.\ 2010, A\&A 520, 54
\bibitem[Beuzit et al.(2004)]{Beuzit2004} Beuzit, J.-L. et al.\ 2004, A\&A 425, 997
\bibitem[Bourke et al.(2006)]{Bourke2006} Bourke, T.L. et al.\ 2006, ApJ 649, L37
\bibitem[Bouy et al.(2003)]{Bouy2003} Bouy, H., Brandner, W., Martin, E., Delfosse, X., Allard, F., \& Basri, G.\ 2003, AJ 126, 1526
\bibitem[Bouy et al.(2004)]{Bouy2004} Bouy, H. et al.\ 2004, A\&A 423, 341
\bibitem[Brandeker et al.(2003)]{Brandeker2003} Brandeker, A., Jayawardhana, R., \& Najita, J. 2003, AJ 126, 2009
\bibitem[Brandeker et al.(2006)]{Brandeker2006} Brandeker, A., Jayawardhana, R., Khavari, P., Haisch, K.E., \& Mardones, D. 2006, ApJ 652, 1572
\bibitem[Burgasser et al.(2003)]{Burgasser2003} Burgasser, A., Kirkpatrick, J.D., Reid, I.N., Brown, M.E., Miskey, C.L., Gizis, J.E. 2003, ApJ, 586, 512
\bibitem[Burgasser et al.(2007)]{Burgasser2007} Burgasser, A., Reid, I.N., Siegler, N., Close, L., Allen, P., Lowrance, P., \& Gizis, J. 2007, in Protostars and Planets V, ed. B. Reipurth, D. Jewitt \& K. Kiel, 427
\bibitem[Burrows et al.(1997)]{Burrows1997} Burrows, A. et al. 1997, ApJ 491, 856
\bibitem[Caballero(2007)]{Caballero2007} Caballero, J.A.\ 2007, ApJ 667, 520
\bibitem[Casewell et al.(2008)]{Casewell2008} Casewell, S.L., Jameson, R.F, \& Burleigh, M.R.\ 2008, MNRAS 390, 1517
\bibitem[Chauvin et al.(2010)]{Chauvin2010} Chauvin, G. et al.\ 2010, A\&A 509, 52
\bibitem[Chabrier et al.(2000)]{Chabrier2000} Chabrier, G., Baraffe, I., Allard, F., \& Hauschildt, P. 2000, ApJ 542, 464
\bibitem[Chen et al.(2006)]{Chen2006} Chen, W.P., Sanchawala, K., \& Chiu, M.C.\ 2006, AJ 131, 990
\bibitem[Christiansen et al.(2008)]{Christiansen2008} Christiansen, J.L. et al.\ 2008, MNRAS 385, 1749
\bibitem[Close et al.(2003)]{Close2003} Close, L.M., Siegler, N., Freed, M., \& Biller, B.\ 2003, ApJ 587, 407
\bibitem[Correia et al.(2006)]{Correia2006} Correia, S., Zinnecker, H., Ratzka, T., \& Sterzik M.F. 2006, A\&A 459, 909
\bibitem[Daemgen et al.(2007)]{Daemgen2007} Daemgen, S., Siegler, N., Reid, I.N., \& Close, L.M. 2007, ApJ 654, 558
\bibitem[Daemgen et al.(2009)]{Daemgen2009} Daemgen, S., Hormuth, F., Brandner, W., Bergfors, C., Janson, M., Hippler, S., \& Henning, T. 2009, A\&A 498, 567
\bibitem[Davis et al.(2010)]{Davis2010} Davis, P.J., Kolb, U. \& Willems, B. 2010, MNRAS 403, 179
\bibitem[Deacon \& Hambly(2007)]{Deacon2007} Deacon, N.R., \& Hambly, N.C. 2007, A\&A 468, 163
\bibitem[Delfosse et al.(2000)]{Delfosse2000} Delfosse, X., Forveille, T., S\'egransan, D., Beuzit, J.-L., Udry, S., Perrier, C., \& Mayor, M. 2000, A\&A 364, 217
\bibitem[Delfosse et al.(2004)]{Delfosse2004} Delfosse, X. et al. 2004, in Spectroscopically and Spatially Resolving the Components of Close Binary Stars, ed. R.W. Hilditch, H. Hensberge \& K. Pavlovski, ASP Conf. Ser. 318, 166
\bibitem[Delorme et al.(2012)]{Delorme2012} Delorme, P., Lagrange, A.M., Chauvin, G., Bonavita, M., Lacour, S., Bonnefoy, M., Ehrenreich, D., \& Beust, H. 2012, A\&A 539, 72
\bibitem[Di\'az et al.(2007)]{Diaz2007} D\'iaz, R.F., Gonz\'alez, J.F., Cincunegui, C., \& Mauas, P.J.D. 2007, A\&A 474, 345
\bibitem[Dommanget \& Nys(2002)]{Dommanget2002} Dommanget, J. \& Nys, O. 2002, VizieR Online Data Catalog, 1274, 0
\bibitem[Dupuy \& Liu(2011)]{Dupuy2011} Dupuy, T.J. \& Liu, M.C. 2011, ApJ 733, 122
\bibitem[Duquennoy \& Mayor(1991)]{Duquennoy1991} Duquennoy, A. \& Mayor, M. 1991, A\&A 248, 485
\bibitem[Eisenhauer et al.(2011)]{Eisenhauer2011} Eisenhauer, F. et al. 2011, The Messenger, 143, 16
\bibitem[Farihi et al.(2005)]{Farihi2005} Farihi, J., Becklin, E.E., \& Zuckerman, B. 2005, ApJS 161, 394
\bibitem[Farihi et al.(2010)]{Farihi2010} Farihi, J., Hoard, D.W., \& Wachter, S. 2010, ApJS 190, 275
\bibitem[Feigelson et al.(2006)]{Feigelson2006} Feigelson, E.D., Lawson, W.A., Stark, M., Townsley, L. \& Garmire, G.P. 2006, AJ 131, 1730
\bibitem[Fischer \& Marcy(1992)]{Fischer1992} Fischer, D. \& Marcy, G. 1992, ApJ 396, 178
\bibitem[Gizis et al.(2000)]{Gizis2000} Gizis, J.E., Monet, D.G., Reid, I.N., Kirkpatrick, J.D., Liebert, J., \& Williams, R.J. 2000, AJ 120, 1085
\bibitem[Gizis et al.(2003)]{Gizis2003} Gizis, J.E., Reid, I.N., Knapp, G.R., Liebert, J., Kirkpatrick, J.D., Koerner, D.W., \& Burgasser, A.J. 2003, AJ 125, 3302
\bibitem[Goodwin et al.(2007)]{Goodwin2007a} Goodwin, S.P., Kroupa, P., Goodman, A., \& Burkert, A. 2007, in Protostars and Planets V, ed. B. Reipurth, D. Jewitt \& K. Kiel, 133
\bibitem[Goodwin \& Whitworth(2007)]{Goodwin2007b} Goodwin, S.P. \& Whitworth, A. 2007, A\&A 466, 943
\bibitem[Gould \& Chanam\'e(2000)]{Gould2004} Gould, A. \& Chanam\'e, J. 2004, ApJS 150, 455
\bibitem[Harrington \& Dahn(1980)]{Harrington1980} Harrington, R.S. \& Dahn, C.C. 1980, AJ 85, 454
\bibitem[Heintz(1992)]{Heinz1992} Heinz, W.D. 1992, ApJS 83, 351
\bibitem[Henry et al.(2006)]{Henry2006} Henry, T.J., Jao, W.-C., Subsavage, J.P., Beaulieu, T.D., Ianna, P.A., Costa, E., \& M\'endez, R.A. 2006, AJ 132, 2360
\bibitem[Hillenbrand \& White(2004)]{Hillenbrand2004} Hillenbrand, L.A. \& White, R.J. 2004, ApJ 604, 741
\bibitem[Hippler et al.(2009)]{Hippler2009} Hippler, S. et al. 2009, The Messenger 137, 14 
\bibitem[Hormuth et al.(2007)]{Hormuth2007} Hormuth, F., Brandner, W., Hippler, S., Janson, M., \& Henning, T. 2007, A\&A 463, 707 
\bibitem[Hormuth et al.(2008)]{Hormuth2008} Hormuth, F., Hippler, S., Brandner, W., Wagner, K., \& Henning, T. 2008, SPIE 7104, 138 
\bibitem[Janson et al.(2007)]{Janson2007} Janson, M., Brandner, W., Lenzen, R., Close, L., Nielsen, E., Hartung, M., Henning, T., Bouy, H. 2007, A\&A 462, 615
\bibitem[Jao et al.(2003)]{Jao2003} Jao, W.-C., Henry, T.J., Subsavage, J.P., Bean, J.L., Costa, E., Ianna, P.A., \& M\'endez, R. 2003, AJ 125, 332
\bibitem[Jenkins et al.(1952)]{Jenkins1952} Jenkins, L.F. 1952, General catalog of trigonometric stellar parallaxes, 1st edition, New Haven, Yale University Observatory
\bibitem[Joergens(2008)]{Joergens2008} Joergens, V. 2008, A\&A 492, 545
\bibitem[Kasper et al.(2007)]{Kasper2007} Kasper, M., Apai, D., Janson, M., \& Brandner, W. 2007, A\&A 472, 321
\bibitem[Kastner et al.(1997)]{Kastner1997} Kastner, J.H., Zuckerman, B., Weintraub, D.A., \& Forveille, T. 1997, Science 277, 67
\bibitem[K\"ohler(2001)]{Kohler2001} K\"ohler, R. 2001, AJ 122, 3325
\bibitem[K\"ohler \& Petr-Gotzens(2002)]{Kohler2002} K\"ohler, R. \& Petr-Gotzens, M.G. 2002, AJ 124, 2899
\bibitem[K\"ohler et al.(2008)]{Kohler2008} K\"ohler, R., Ratzka, T., Herbst, T.M., \& Kasper, M. 2008, A\&A 482, 929
\bibitem[Konopacky et al.(2007)]{Konopacky2007} Konopacky, Q.M., Ghez, A., Duchene, G., McCabe, C., \& Macintosh, B.A. 2007, AJ 133, 2008
\bibitem[Konopacky et al.(2010)]{Konopacky2010} Konopacky, Q.M., Ghez, A., Barman, T.S., Rice, E.L., Bailey, J.I., White, R.J. \& Duchene, G. 2010, ApJ 711, 1087
\bibitem[Kouwenhoven et al.(2007)]{Kouwenhoven2007} Kouwenhoven, M.B.N., Brown, A.G.A., Portegies Zwart, S.F., \& Kaper, L. 2007, A\&A 474, 77
\bibitem[Kraus \& Hillenbrand(2007)]{Kraus2007} Kraus, A.L. \& Hillenbrand, L.A. 2007, AJ 134, 2340
\bibitem[Law et al.(2006)]{Law2006} Law, N.M., Hodgkin, S.T, \& Mackay, C.D.\ 2006, MNRAS 368, 1917
\bibitem[Law(2006)]{Law2006thesis} Law, N.M.\ 2006, Ph.D. thesis, Institute of Astronomy \& Selwyn College, Cambridge University
\bibitem[Law et al.(2008)]{Law2008} Law, N.M., Hodgkin, S.T, \& Mackay, C.D.\ 2008, MNRAS 384, 150
\bibitem[L\'epine \& Bongiorno(2007)]{Lepine2007} L\'epine, S. \& Bongiorno, B.\ 2007, AJ 133, 889
\bibitem[L\'epine et al.(2009)]{Lepine2009} L\'epine, S., Thorstensen, J.R., Shara, M.M., \& Rich, R.M.\ 2009, AJ 137, 3632
\bibitem[L\'epine \& Gaidos(2011)]{Lepine2011} L\'epine, S. \& Gaidos, E.\ 2011, AJ 142, 138
\bibitem[Li \& Hu(1998)]{Li1998} Li, J.Z. \& Hu, J.Y.\ 1998, A\&AS 132, 173
\bibitem[Li et al.(2000)]{Li2000} Li, J.Z., Hu, J.Y., \& Chen, W.P.\ 2000, A\&A 356, 157
\bibitem[Liu et al.(2008)]{Liu2008} Liu, M.C., Dupuy, T.J., \& Ireland, M.J.\ 2008, ApJ 689, 436
\bibitem[L\'opez-Santiago et al.(2006)]{Lopez2006} L\'opez-Santiago, J., Montes, D., Crespo-Chac\'on, I, \& Fern\'andez-Figueroa, M.J.\ 2006, ApJ 643, 1160
\bibitem[Lowrance et al.(1999)]{Lowrance1999} Lowrance, P.J. et al.\ 1999, ApJ 512, L69
\bibitem[Lowrance et al.(2005)]{Lowrance2005} Lowrance, P.J. et al.\ 2005, AJ 130, 1845
\bibitem[Luhman et al.(2005)]{Luhman2005} Luhman, K.L. et al.\ 2005, ApJ 631, 69
\bibitem[Macintosh et al.(2001)]{Macintosh2001} Macintosh, B. et al.\ 2001, in ASP Conf. Ser. 244, Young stars near the Earth: Progress and Prospects, eds. R. Jayawardhana \& T.P. Greene, San Fransisco, 309
\bibitem[Mamajek(2005)]{Mamajek2005} Mamajek, E.E. 2005, ApJ 634, 1385
\bibitem[Mason et al.(2000)]{Mason2000} Mason, B.D., Wycoff, G.L., Urban, S.E., Hartkopf, W.I., Holdenreid, E.R., \& Makarov, V.V. 2000, AJ 120, 3244
\bibitem[Mason et al.(2001)]{Mason2001} Mason, B.D., Wycoff, G.L., Hartkopf, W.I., Douglass, G.G., \& Worley, C.E. 2001, AJ 122, 3466
\bibitem[Mason et al.(2009)]{Mason2009} Mason, B.D., Hartkopf, W.I., Gies, D.R., Henry, T.J., \& Helsel, J.W. 2009, AJ 137, 3358
\bibitem[McCarthy et al.(2001)]{McCarthy2001} McCarthy, C., Zuckerman, B., \& Becklin, E.E. 2001, AJ 121, 3259
\bibitem[Morlet et al.(2002)]{Morlet2002} Morlet, G., Salaman, M., \& Gili, R. 2002, A\&A 396, 933
\bibitem[Neuh\"auser et al.(2000)]{Neuhauser2000} Neuh\"auser, R. et al.\ 2000, A\&A 354, 9
\bibitem[Norton et al.(2007)]{Norton2007} Norton, A.J. et al. 2007, A\&A 467, 785
\bibitem[Parihar et al.(2009)]{Parihar2009} Parihar, P., Messina, S., Bama, P., Medhi, B.J., Muneer, S., Velu, C., \& Ahmad, A. 2009, MNRAS 395, 593
\bibitem[Perryman et al.(1997)]{Perryman1997} Perryman, M.A.C. et al.\ 1997, A\&A  323, 49
\bibitem[Peter et al.(2012)]{Peter2012} Peter, D., Feldt, M., Henning, T., \& Hormuth, F.\ 2012, A\&A  538, 74
\bibitem[Quirrenbach et al.(2010)]{Quirrenbach2010} Quirrenbach, A. et al. 2010, SPIE 7735, 37
\bibitem[Raghavan et al.(2010)]{Raghavan2010} Raghavan, D. et al.\ 2010, ApJS 190, 1
\bibitem[Reid \& Cruz(2002)]{Reid2002} Reid, I.N., \& Cruz, K.L.\ 2002, AJ 123, 2806
\bibitem[Reid et al.(2004)]{Reid2004} Reid, I.N. et al.\ 2004, AJ 128, 463
\bibitem[Reid et al.(2007)]{Reid2007} Reid, I.N., Cruz, K.L., \& Allen, P.R.\ 2007, AJ 133, 2825
\bibitem[Reidel et al.(2010)]{Reidel2010} Reidel, A.R. et al.\ 2010, AJ 140, 897
\bibitem[Reipurth \& Clarke(2001)]{Reipurth2001} Reipurth, B. \& Clarke, C.\ 2001, AJ 122, 432
\bibitem[Riaz et al.(2006)]{Riaz2006} Riaz, B., Gizis, J., \& Harvin, J. 2006, AJ 132, 866
\bibitem[Riaz et al.(2012)]{Riaz2012} Riaz, B., Lodieu, N., Goodwin, S., Stamatellos, D., \& Thompson, M. 2012, MNRAS 420, 2497
\bibitem[R\"oser et al.(2010)]{Roeser2010} R\"oser, S., Demleitner, M., \& Schilbach, E. 2010, AJ 139, 2440
\bibitem[Salim \& Gould(2003)]{Salim2003} Salim, S. \& Gould, A. 2003, ApJ 582, 1011
\bibitem[S\'egransan et al.(2000)]{Segransan2000} S\'egransan, D., Delfosse, X., Forveille, T., Beuzit, J.-L., Udry, S., Perrier, C., \& Mayor, M. 2000, A\&A 364, 665
\bibitem[Seymour et al.(2002)]{Seymour2002} Seymour, D.M., Mason, B.D., Hartkopf, W.I., \& Wycoff, G.L. 2002, AJ 123, 1023
\bibitem[Sicilia-Aguilar et al.(2008)]{Sicilia-Aguilar2008} Sicilia-Aguilar, A. et al.\ 2008, ApJ 673, 382
\bibitem[Shatsky \& Tokovinin(2002)]{Shatsky2002} Shatsky, N. \& Tokovinin, A. 2002, A\&A 382, 92
\bibitem[Shkolnik et al.(2010)]{Shkolnik2010} Shkolnik, E., Hebb, L., Liu, M., Reid, N., \& Cameron, A.C. 2010, ApJ 716, 1522
\bibitem[da Silva et al.(2009)]{daSilva2009} da Silva, L., Torres, C.A.O., de La Reza, R., Quast, G.R., Melo, C.H.F., \& Sterzik, M.F. 2009, A\&A 508, 833
\bibitem[Skrutskie et al.(2006)]{Skrutskie2006} Skrutskie, M.F. et al.\ 2006, AJ 131, 1163
\bibitem[Smart et al.(2010)]{Smart2010} Smart, K.J., Ioannidis, G., Jones, H.R.A, Bucciarelli, B., \& Lattanzi, M.G.\ 2010, A\&A 511, 30
\bibitem[Stassun et al.(1999)]{Stassun1999} Stassun, K.G., Mathieu, R.D., Mazeh, T., \& Vrba, F.J.\ 1999, AJ 117, 2941
\bibitem[Strigachev et al.(2004)]{Strigachev2004} Strigachev, A. \& Lampens, P. 2004, A\&A 422, 1023
\bibitem[Thies \& Kroupa(2007)]{Thies2007} Thies, I. \& Kroupa. P. 2007, ApJ 671, 767
\bibitem[Tokovinin et al.(2006)]{Tokovinin2006} Tokovinin, A., Thomas, S., Sterzik, M., \& Udry, S. 2006, A\&A 450, 681
\bibitem[Torres et al.(2002)]{Torres2002} Torres, G., Neuh\"auser, R., \& Guenther, E. 2002, AJ 123, 1701
\bibitem[Tubbs et al.(2002)]{Tubbs2002} Tubbs, R.N., Baldwin, J.E., Mackay, C.D., \& Cox, G.C. 2002, A\&A 387, 21
\bibitem[van Altena et al.(1995)]{vanAltena1995} van Altena, W.F., Lee, J.T., \& Hoffleit, E.D. 1995, The general catalog of trigonometric stellar parallaxes, eds. W.F van Altena, J.T. Lee \& E.D. Hoffleit
\bibitem[van der Marel et al.(2002)]{Marel2002} van der Marel, R.P., Gerssen, J., Guhathakurta, P., Peterson, R.C., \& Gebhardt, K. 2002, AJ 124, 3255
\bibitem[Vennes et al.(1999)]{Vennes1999} Vennes, S., Thorstensen, J.R., \& Polomski, E.F. 1999, ApJ 523, 386
\bibitem[Voges et al.(1999)]{Voges1999} Voges, W. et al.\ 1999, A\&A 349, 389
\bibitem[Webb et al.(1999)]{Webb1999} Webb, R.A., Zuckerman, B., Platais, I., Patience, J., White, R.J., Schwartz, M.J., \& McCarthy, C. 1999, ApJ 512, L63
\bibitem[Weis(1991)]{Weis1991} Weis, E. 1991, AJ 101, 1882
\bibitem[Zapatero Osorio et al.(2004)]{Zapatero2004} Zapatero Osorio, M.R., Lane, B.F., Pavlenko, Y., Mart\'in, E.L., Britton, M., Kulkarni, S.R. 2004, ApJ 615, 958
\bibitem[Zuckerman et al.(2001)]{Zuckerman2001} Zuckerman, B., Song, I., Bessell, M.S. \& Webb, R.A. 2001, ApJ 562, L87

\end{thebibliography}
\end{document}